\documentclass[%
reprint,
superscriptaddress,
amsmath,amssymb,
aps,
floatfix,
]{revtex4-2}

\usepackage[
activate={true,compatibility},
final,
tracking=true,
kerning=true,
nopatch=footnote]{microtype}
\usepackage{mathtools}
\usepackage{amsfonts}
\usepackage{nicefrac}
\usepackage{physics}
\usepackage{color}
\usepackage{graphicx}%
\usepackage{dcolumn}%
\usepackage{hyperref} %
\usepackage{booktabs}
\usepackage{tabularx}
\usepackage{multirow}
\usepackage{subcaption}
\usepackage{xcolor}
\usepackage[normalem]{ulem} %
\usepackage[noend]{algpseudocode}
\usepackage{amsthm}
\usepackage{algorithm}     
\usepackage{tikz}
\usetikzlibrary{positioning}        %
\usepackage[margin=1in]{geometry} 

\DeclareRobustCommand{\rchi}{{\mathpalette\irchi\relax}}
\newcommand{\irchi}[2]{\raisebox{\depth}{$#1\chi$}} %

\makeatletter
\renewcommand\@makecaption[2]{%
  \par
  \vskip\abovecaptionskip
  \begingroup
   \small\rmfamily
    \begingroup
     \samepage
     \flushing
     \let\footnote\@footnotemark@gobble
     \@make@capt@title{#1}{#2}\par
    \endgroup
  \endgroup
  \vskip\belowcaptionskip
}
\makeatother

\newcolumntype{M}{>{\centering}X}
\newcolumntype{Y}{>{\hsize=.17\textwidth\arraybackslash}X}
\newcolumntype{C}{>{\hsize=.115\textwidth\centering\arraybackslash}X}
\newcolumntype{R}{>{\hsize=.115\textwidth\raggedleft\arraybackslash}X}

\begin{document}

\keywords{tensor networks;turbulence;fluid dynamics;quantum-inspired methods; numerical solver; matrix product states;tensor train}

\title{Compression, simulation, and synthesis of turbulent flows with tensor trains}

\author{Stefano \surname{Pisoni}}
\affiliation{Quantum Research Center, Technology Innovation Institute, Abu Dhabi, UAE}
\affiliation{Institute for Quantum-Inspired and Quantum Optimization, Hamburg University of Technology, Germany}
\author{Raghavendra D. \surname{Peddinti}}
\affiliation{Quantum Research Center, Technology Innovation Institute, Abu Dhabi, UAE}
\author{Siddhartha E. \surname{Guzman}}
\affiliation{Quantum Research Center, Technology Innovation Institute, Abu Dhabi, UAE}
\author{Egor \surname{Tiunov}}
\affiliation{Quantum Research Center, Technology Innovation Institute, Abu Dhabi, UAE}
\author{Leandro \surname{Aolita}}
\affiliation{Quantum Research Center, Technology Innovation Institute, Abu Dhabi, UAE}

\date{\today}

\begin{abstract}
Numerical simulations of turbulent fluids are paramount to real-life applications, from predicting and modeling flows to diagnostic purposes in engineering. However, they are also computationally challenging due to their intrinsically non-linear dynamics, which require a very high spatial resolution to accurately describe them.
A promising idea is to represent flows on a discrete mesh using tensor trains (TTs), featuring a convenient scaling of the number of parameters with the mesh size.
However, it is unclear how the compression power of TTs is affected by the complexity of the flows, as measured by the Reynolds number. 
In fact, no comprehensive analysis of how the TT representation affects the turbulent properties has yet been carried out.
We fill this gap by analyzing TTs as an Ansatz to compress, simulate, and generate 3D snapshots with turbulent-like features. Specifically, we first investigate the effect of TT compression on key turbulence signatures, such as the energy spectrum, the PDF of velocity increments, and flatness. Second, we extend the 2D TT-solver introduced in~\cite{peddinti} to a 3D cubic domain with periodic boundary conditions. We use it to simulate the incompressible Navier-Stokes dynamics at $Re_{\lambda}=315$ for a total of 9-10 Kolmogorov turnover times, showcasing the numerical stability of the TT-solver in fully developed turbulent regimes.
Third, we develop a TT algorithm to synthesize artificial snapshots that exhibit turbulent-like features, with a logarithmic cost in the mesh size.
Our analysis demonstrates the ability of the TT representation to capture the characteristic features of turbulence.
This offers a powerful quantum-inspired toolkit for the computational treatment of turbulent flows.

\end{abstract}

\maketitle

\section{Introduction} \label{section:Introduction} 

Understanding turbulence is a long-standing open problem in classical physics. Computational fluid dynamics (CFD) plays a key role in that quest, as it aims at numerically simulating turbulent flows. 
However, in standard numerical approaches, such as direct numerical simulations (DNS), accuracy is often hindered by the large mesh sizes required to capture the multiscale properties of turbulent flows~\cite{pope_2000}.
This is the CFD incarnation of the infamous curse of dimensionality that affects many computational problems. In this regard, tensor networks (TNs) have emerged as a new computational tool that has proven crucial in broadening the set of problems that can be tackled with numerical techniques, from machine learning models~\cite{han2018unsupervised} to the simulation of many quantum systems~\cite{verstraete2008matrix,orus2014practical}.
TNs enable efficient data compression of high-dimensional objects while still preserving their fundamental features and correlations.
The best-known example is the celebrated matrix product state (MPS)~\cite{verstraete2008matrix,orus2014practical}, also referred to as tensor train (TT)~\cite{oseledets2011tensortrain,Khoromskij_2011}. This TN was originally proposed for 1D quantum systems~\cite{White_PRL_1992}.
However, its versatility and ease of manipulation have allowed it to adapt well to diverse scenarios~\cite{Image_Compression,option_pricing}, recently including fluid simulations~\cite{gourianov2022quantum,peddinti,peddinti2025technicalreportquantuminspiredsolver,kiffner_jaksch2023tensor,kornev2023chemicalmixer, arenstein2025fast, Turbulent_pdf_tensor_networks_Gourianov, quantuminspiredtensornetworkfractionalstepmethod}.

Indeed, the seminal work~\cite{gourianov2022quantum} opened a research program aimed at simulating the time evolution of a fluid field within the compressed TT representation. 
To this end, significant efforts have been put into translating standard CFD practices to TT native methods, from the treatment of non-trivial boundary conditions, including immersed bodies with complex geometries, to numerical schemes for the efficient retrieval of the solution from its TT representation~\cite{peddinti,kiffner_jaksch2023tensor, kornev2023chemicalmixer}.
These TT-based CFD solvers have reported promising results, indicating computational advantages both in memory and runtime.
The general intuition is that the scale separation induced by the Kolmogorov energy cascade mechanism~\cite{kolmogorov1941local}, whereby energy is transferred locally from one spatial scale to the next smaller one, should render TT representations efficient, in analogy with local interactions in 1D quantum systems.
However, the computational resources required by these novel algorithms for addressing turbulent flow scenarios have not yet been systematically investigated.
Specifically, the investigation of the key statistical metrics characterizing turbulence within the context of tensor-train (TT) solvers remains absent.
This is particularly unsettling, since turbulent flows precisely define the regime in which the computational advantages potentially offered by TTs are needed the most.

With this motivation, we conduct a comprehensive analysis to benchmark the TT encoding for turbulent flows.
Throughout the work, we consider incompressible fluids in a periodic cubic domain and examine the statistical properties of turbulent velocity fields---such as the energy spectrum, flatness, and PDF of velocity increments---as our main metrics.
We divide our paper into three aspects:
first, we consider \textit{single snapshot compression}. We encode turbulent snapshots on a 3D mesh of $1024^3=2^{30}$ pixels into TTs of $N=30$ tensors (i.e., 30-qubit MPSs). 
We check the statistical metrics of the resulting TTs when varying the bond dimension $\chi$ (i.e., the maximal tensor size), which controls the compression rate of the TT encoding.
The snapshots are taken from a reference DNS data set~\cite{Cardesa_2017}, with $Re_{\lambda}=315$.
We show that the TT with $\chi=1000$, which retains the $2.2\%$ of the total number of parameters, already reproduces the statistical metrics under consideration with small deviations from the ground truth. We also find that the ordering of the matrices in the TT is a crucial choice for this task, with the \textit{interleaved} one outperforming the \textit{stacked} one.
To our knowledge, this constitutes the most rigorous analysis of TT compression of a fully developed turbulent flow.
Second, we present a \textit{TT-based 3D fluid solver} whose memory and runtime scale polylogarithmically with the mesh size. This is an extension of the solver presented in~\cite{peddinti}. 
We initialize the flow on a snapshot taken from the data set mentioned above~\cite{Cardesa_2017}.
The time evolution is performed with $\chi=100$ (retaining only the $0.03\%$ of the total parameters), in a homogeneous and isotropic setting with $Re_{\lambda}=315$, a regime that has never been tested before using TT-based solvers.
The physical evolution time covers 9-10 Kolmogorov turnover times.
Third, we develop an efficient TT algorithm to generate \textit{synthetic snapshots} that exhibit some features of turbulent flows. The runtime of this algorithm scales polynomially with the bond dimension and linearly with the number of spatial scales.
It relies on a novel interpolation scheme for TT-encoded fields, which we introduce as a technical contribution and that is potentially interesting beyond the current scope~\cite{siddhartha_2025}.

\section{Preliminaries}\label{section:preliminaries}

\subsection{TT formalism}\label{subsection:QTT formalism}

We introduce the tensor train (TT) formalism---also known as matrix product state (MPS)~\cite{White_PRL_1992,oseledets2011tensortrain}---directly applied to the encoding of the velocity vector field $\boldsymbol{v}(\boldsymbol{x}) = (u(\boldsymbol{x}),v(\boldsymbol{x}),w(\boldsymbol{x}))$. Each velocity component is a scalar field discretized on a $1024^3$ grid and is represented as an individual TT. Then the 3D domain constitutes a mesh of $2^N$ points, specified by $N=N_x+N_y+N_z=30$ bits, which corresponds to the total number of tensors in the TT. 
For instance, the velocity component $u(x,y,z)$ (the same holds true for $v$ and $w$) is given by the vector of elements $u_{\textbf{i},\textbf{j}, \textbf{k}} := u(x_{\textbf{i}},y_{\textbf{j}},z_{\textbf{k}})$, where the binary strings ${\textbf{i}}:=(i_1,i_2, \hdots, i_{Nx})$, ${\textbf{j}}:=(j_1,j_2, \hdots, j_{Ny})$ and ${\textbf{k}}:=(k_1,k_2, \hdots, k_{Nz})$ index the discretized coordinates $x_{\textbf{i}}$, $y_{\textbf{j}}$ and $z_{\textbf{k}}$, respectively.

Then, each $u_{\textbf{i} ,\textbf{j}, \textbf{k}}$ is a product of $N$ matrices:
\begin{multline}
\label{eq:MPS_sta}
u_{\textbf{i},\textbf{j}, \textbf{k}} = A_1^{(i_1)} A_2^{(i_2)}\dots A_{N_x}^{(i_{N_x})}\,B_1^{(j_1)} B_2^{(j_2)}\dots \\
\dots B_{N_y}^{(j_{N_y})}\,C_1^{(k_1)} C_2^{(k_2)}\dots C_{N_z}^{(k_{N_z})}.
\end{multline}
This is the TT (or MPS) representation. When the indices $i_l$, $j_l$, and $k_l$---referred to as \textit{physical indices}---are binary valued like in our case, i.e. $i_l, j_l, k_l \in \{0,1\}$, Eq.~\eqref{eq:MPS_sta} is sometimes referred to as quantics TT (QTT).
Then the bond dimension $\chi$ is defined as the maximum dimension over all $2N$ matrices used.
Importantly, the total number of parameters is at most $2N\chi^2$. Hence, the TT representation provides an exponential compression of the $2^N$-dimensional vector when $\chi$ is constant or scales polynomially with $N$.

\begin{figure}[t!]
    \centering
    \includegraphics[width=\linewidth]{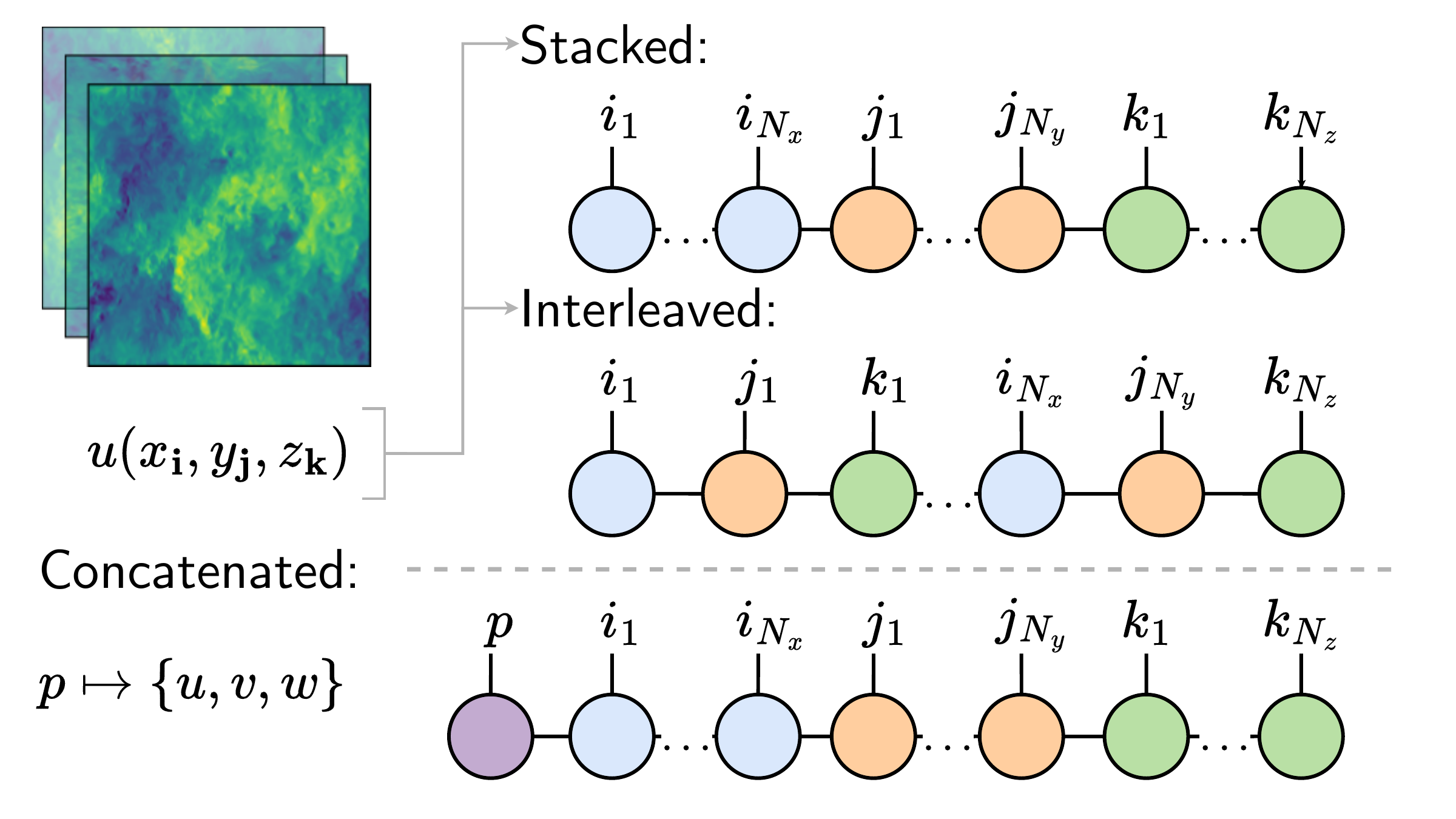}
    \caption{
    \textbf{Different TT encodings of a velocity field.} The discretized velocity component $u(x_{\textbf{i}},y_{\textbf{j}},z_{\textbf{k}})$ can be decomposed into two different types of TTs: stacked and interleaved. In both cases, the TT consists of a 1D chain of tensors connected over their virtual (horizontal) indices. Each tensor has a physical (vertical) index labeled by a bit $i_l$, $j_l$, or $k_l$ in the binary representation of  $(\textbf{i},\textbf{j},\textbf{k})$. These binary indices naturally define a notion of spatial scales, with $(i_m,j_m,k_m)$ signifying the $m$-$th$ subdivision of the dyadic grid.
    The maximal cardinality over all virtual indices is called the \emph{bond dimension} of the TT, which captures the amount of inter-scale correlations. The stacked and interleaved encodings differ in the ordering of the binary indices, as described in Eqs.~\ref{eq:MPS_sta} and~\ref{eq:MPS_inter}. The three components of the velocity field ($u,v,w$) are encoded into three individual TTs. These can, in turn, be represented by a single TT using an additional tensor with a 3-dimensional physical index $p$, defined as the concatenated TT representation of the full velocity field. An example of this is shown at the bottom for the stacked encoding.}
    \label{fig:encodings}
\end{figure}

Eq.~\eqref{eq:MPS_sta} is often referred to as the \textit{stacked encoding} because the matrices are ordered according to the physical dimensions of the 3D grid. However, other arrangements are possible. In particular, we will also consider the following arrangement:

\begin{multline}
\label{eq:MPS_inter}
    u_{\textbf{i},\textbf{j}, \textbf{k}} = A_1^{(i_1)}B_1^{(j_1)}C_1^{(k_1)}A_2^{(i_2)}B_2^{(j_2)}C_2^{(k_2)}\dots \\ 
    \dots A_{N_x}^{(i_{N_x})}B_{N_y}^{(j_{N_y})}C_{N_z}^{(k_{N_z})}.
\end{multline}
This is known as the \textit{interleaved} encoding. Fig.~\ref{fig:encodings} depicts diagrammatically the two encodings as well as the additional \textit{concatenated} encoding, which is used in Appendix~\ref{app:Concatenated encoding}.

We remark that the binary discretization of the domain naturally defines a notion of spatial scales. 
For example, in a 1D domain, each TT tensor $A_l^{(i_l)}$ labels the discretization at the scale $l$ in a dyadic fashion. 
Therefore, the chosen arrangement of tensors defines the correlation structure across the spatial scales of the domain, and finding the optimal arrangement amounts to minimizing inter-scale correlations, leading to a TT encoding with the fewest number of parameters.
From a quantum information point of view, the correlations embedded in the state, defined in terms of the entanglement entropy, determine the matrix dimensions in the TT representation. In particular, having local interactions in 1D quantum systems was shown to be a necessary condition for the corresponding ground states to be well approximated by a TT with low $\chi$~\cite{schollwock2011density,orus2014practical}.
This suggests that turbulent signals, which exhibit stronger couplings for neighboring scales as we will see in Sec.~\ref{subsection:Turbulence}, might also be well captured by TT with low $\chi$s.

\begin{figure*}[t]
    \centering
     \includegraphics[width=\textwidth]{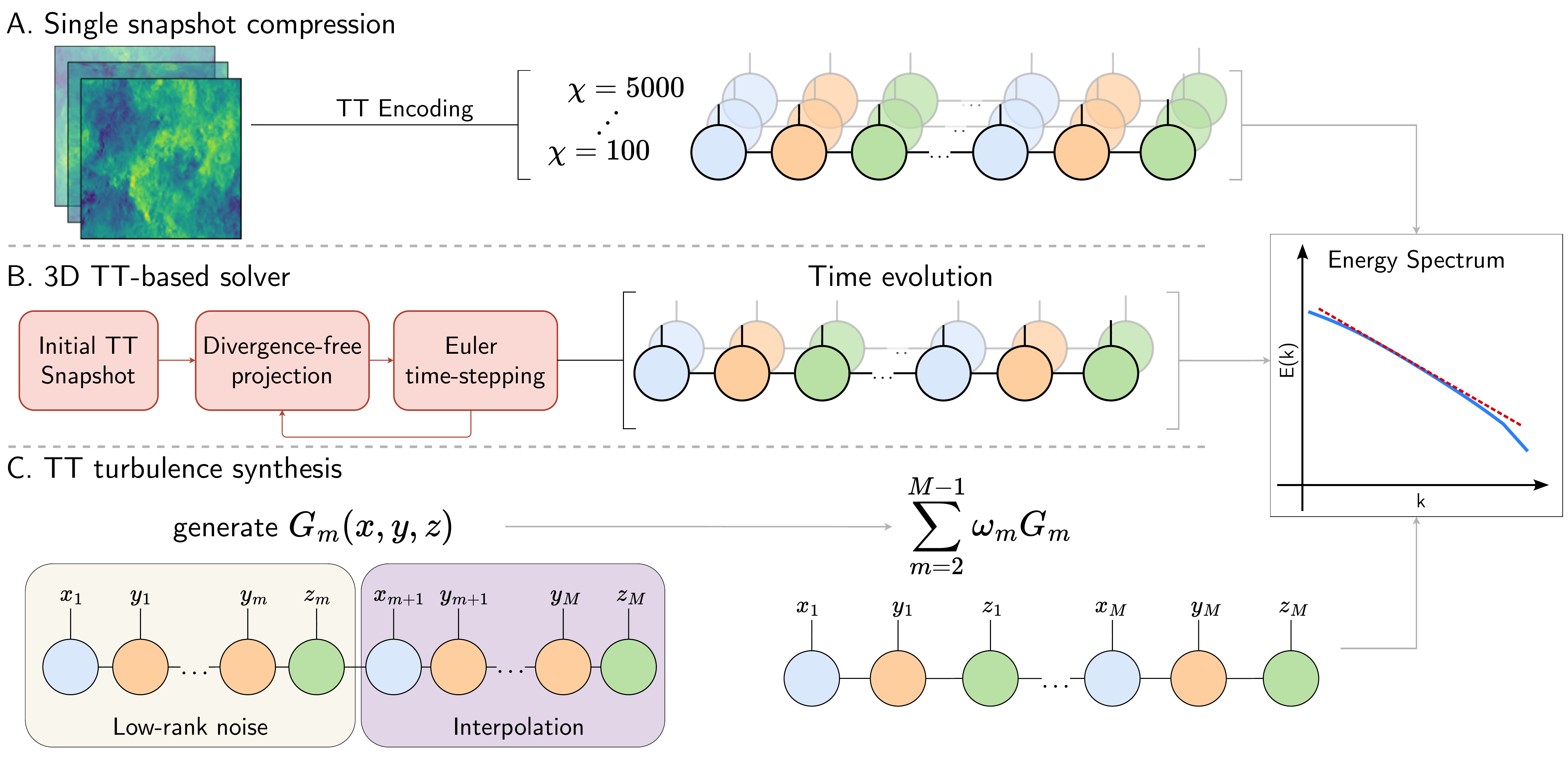}
     \caption{
     \textbf{Schematic representation of the three investigated aspects of turbulence.}
     We numerically investigate three different settings to analyze the TT encoding of turbulent flows.
     \textbf{A. Single snapshot compression.} Here, we encode the velocity field at a given time, called \textit{a snapshot}, into its corresponding TT representation. We use the well-known turbulence DNS dataset~\cite{Cardesa_2017} and compare key statistical metrics of the compressed TT snapshots with those of the original dataset for increasing bond dimensions $\chi$. We perform this analysis for both stacked and interleaved encoding. The results are reported in Sec.~\ref{subsection:Single snapshot compression}.
     \textbf{B. 3D TT-based solver.} Using the TT snapshot obtained from the DNS solution as the initial condition, we simulate the time evolution of the flow completely within the TT representation. We first project onto the divergence-free manifold of the velocity field and perform the time stepping using an explicit Euler scheme. We compute the energy spectrum of the time-series of solutions obtained. We report the results for the interleaved encoding in Sec.~\ref{subsection:3D MPS-based solver}.
     \textbf{C. TT snapshots synthesis.}
     Here, we construct a TT field that exhibits some key turbulent features: the divergence-free condition, the Kolmogorov energy spectrum, and intermittency, quantified by the flatness. 
     The algorithm generates random low-rank TTs ($\chi=10$) at each spatial scale $m$, then interpolates them to the desired resolution $M$.
     The final snapshot is the summation of these TTs weighted by the appropriate weights $\omega_m$. We compute and verify these properties for an ensemble of 20 snapshots. In this instance, we restrict ourselves to the interleaved encoding. The detailed explanation with results is outlined in Sec.~\ref{subsection:Synthetic_turbulence}.
     }\label{fig:diagram}
\end{figure*}

\subsection{Turbulence}\label{subsection:Turbulence}
Here, we introduce the key concepts in the study of turbulent fields. First, we recap the theory of homogeneous and isotropic turbulence (HIT)~\cite{benzi_homogeneous_2015} and clarify why this setting is well-suited for the TT framework. We will later use these results to benchmark our numerical investigations in Sec.~\ref{section:Numerical results}. Second, we present the key challenges with the current approaches to turbulence simulation.

\paragraph{Isotropic and homogeneous turbulence:}\label{subsection:HIT}
The word turbulence usually refers to a chaotic and multi-scale behavior of fluid flows in space and time. The complexity of turbulent flows stems from the Navier-Stokes equations:
\begin{subequations}\label{eq:NS}
\begin{align} 
        \frac{\partial{\boldsymbol{v}}}{\partial{t}} + (\boldsymbol{v} \cdot \nabla )\, \boldsymbol{v} &= -\frac{1}{\varrho} \nabla p + \nu\, \nabla^2\, \boldsymbol{v}, \label{eq:NS_eq_a} \\
        \nabla \cdot \boldsymbol{v} &= 0, \label{eq:NS_eq_b}
\end{align}
\end{subequations}
where $\boldsymbol{v}=\boldsymbol{v}(\boldsymbol{x},t)$ and $p=p(\boldsymbol{x},t)$ are respectively the velocity and pressure fields at position $\boldsymbol{x}$ and time $t$, $\varrho$ is the  density, and $\nu$ the kinematic viscosity. Eqs.~\eqref{eq:NS_eq_a} and~\eqref{eq:NS_eq_b} follow respectively from momentum and mass conservation~\cite{pope_2000}. In this work, only 3D incompressible fluids are considered, a condition enforced by Eq.~\eqref{eq:NS_eq_b}. 
The emergence of a turbulent phase is associated with a scalar parameter, the Reynolds number ($\mathrm{Re}$), which describes the degree of turbulence in a flow. Specifically, denoting by $v_0$ the characteristic scale of velocity fluctuations and by $L$ the scale characterizing energy input, $\mathrm{Re} \coloneqq v_0 L / \nu$.

Given the chaotic nature of turbulence, a statistical approach has been developed through the years~\cite{monin2007,monin2013statistical}. In particular, being a system strongly out of equilibrium, new tools have been developed, starting from the observation that the energy dissipation $\epsilon = \nu (\langle \nabla \boldsymbol{v})^2 \rangle$ is independent of the Reynolds number, where $\langle \dots \rangle$ means an average in space and time.
Specifically, $\epsilon \sim const.$ as $\mathrm{Re} \rightarrow \infty$.

In 1941~\cite{kolmogorov1941local}, Kolmogorov clearly highlighted this fundamental feature of turbulence and showed that, in the homogeneous and isotropic turbulence (HIT) setting, two separated ranges of scales emerge: the inertial range and the dissipative range, separated by the Kolmogorov scale $\eta = (\nu^3 / \epsilon)^{1/4}$. 
Specifically, introducing the longitudinal velocity fluctuations $\delta v (r) \coloneqq (\boldsymbol{v}(\boldsymbol{x}+\boldsymbol{r})-\boldsymbol{v}(\boldsymbol{x})) \cdot \boldsymbol{r} / r$, it was shown that in the inertial range ($r>>\eta$) these are solely controlled by $\epsilon$ and $r$, while in the dissipative range ($r \sim \eta$) the dissipative forces become important.
Importantly, the induced scale separation implies that what happens at the tiny scales must be independent of what happens at the large scales, suggesting that the interactions among them decay with increasing scale separation. This was indeed confirmed by analyzing the non-linear advection term in momentum space~\cite{EYINK_locality_turbulent_cascade, Kraichnan_1959_HIT_high_Reynolds, biferale2003shell}.
This local interaction among scales is what ultimately motivates our study because it corresponds to a local correlation among the tensors in the TT representation, which reflects in an efficient encoding as outlined in Sec.~\ref{subsection:Turbulence}.
Moreover, by assuming that the statistical properties of turbulent flows are scale invariant in the inertial range and using only dimensional arguments, one can further conclude that:
\begin{equation} \label{eq:vel_fluc_p}
    S_p(r) = \langle \delta v (r)^p \rangle \sim \epsilon^{p/3} r^{p/3}.
\end{equation}
Eq.~\eqref{eq:vel_fluc_p} gives rise to the turbulent kinetic energy spectrum, whose Fourier representation in the inertial range follows the celebrated power law:
\begin{equation}\label{eq:E_k}
    E(k) \propto \epsilon^{2/3} k^{-5/3} ,
\end{equation}
where $k = |\vec{k}|$ is the wave number magnitude. This result will be used in Sec.~\ref{section:Numerical results} as a benchmark for our numerical results.
We remark that Eq.~\eqref{eq:vel_fluc_p} requires scale-invariance to hold, which is an assumption of Kolmogorov theory. Therefore, deviations from Eq.~\eqref{eq:vel_fluc_p} and Eq.~\eqref{eq:E_k} might be observed, giving rise to intermittent phenomena.
Intermittency is another fundamental feature of turbulence that is associated with regions of high vorticity in the flow. One standard way to evaluate it is via the so-called flatness, or generalized kurtosis, defined as:
\begin{equation} \label{eq:flatness}
    \Gamma_p(r) \equiv S_p(r) / S_2(r)^{p/2}.
\end{equation}
Note that Eq.~\eqref{eq:vel_fluc_p} predicts $\Gamma_p(r) \sim const$ in the inertial range. However, $\Gamma_p(r)$ is observed to increase for $r \rightarrow \eta$, both in DNS and laboratory experiments~\cite{universal_stats_turb_experiments_2008}.
Therefore, we also look at the flatness as an additional metric to benchmark our numerical results in Sec.~\ref{section:Numerical results}.
Lastly, we remark that the intermittent nature of a turbulent signal can also be inferred from the probability distribution function (PDF) of the longitudinal velocity increments. In fact, the emergence of a non-Gaussian statistics, characterized by fat tails signaling the occurrence of extreme events, marks the turbulent phase of a fluid flow~\cite{benzi_homogeneous_2015}. We also use this metric to benchmark our numerical results in Sec.~\ref{section:Numerical results}.

\paragraph{Turbulence simulation:}\label{subsection:Turbulence simulation}
Simulating turbulent flows has always been a major challenge in computational physics.
Ideally, one would simulate any flow regime using direct numerical simulation (DNS), a mesh-based method that evolves the discretized version of Eqs.~\eqref{eq:NS} in time without additional modeling.
However, for turbulent flows, the required mesh size for an accurate DNS solution increases with the Reynolds number. 
Specifically, the minimal number of spatial scales $M$ to be resolved is given by $M \sim \frac{3}{4} \log \mathrm{Re}$. 
In fact, as discussed in Sec.~\ref{subsection:Turbulence}, turbulent fluctuations span a large range of scales, from the Kolmogorov scale $\eta$ associated with the tiniest whirls of turbulence to the integral scale $L$ where energy is injected into the system. 
This is referred to as the inertial range, and it is well known that $\frac{\eta}{L} \sim \mathrm{Re}^{-3/4}$.
This scale separation is what ultimately hinders DNS of turbulent flows.
From the TT perspective, since each tensor in the TT represents a scale, at least $M$ tensors are needed per spatial dimension. 
\section{Results}\label{section:Numerical results}

In this section, we present the numerical results of this work.
First, in Sec.~\ref{subsection:Single snapshot compression}, we compress turbulent snapshots and look at how the key statistical measures (introduced in Sec.\ref{subsection:Turbulence}) are reproduced when varying the number of parameters in the TT representation, controlled by $\chi$. 
Next, in Sec.~\ref{subsection:3D MPS-based solver}, we use turbulent snapshots to initialize and benchmark the time evolution performed with our 3D TT-based solver. 
Finally, in Sec.~\ref{subsection:Synthetic_turbulence} we present an efficient algorithm to synthesize snapshots with turbulent-like features in the TT representation.

For the first two sub-sections, we use the DNS of isotropic turbulence generated in Minotauro at the Barcelona Supercomputing Center~\cite{Cardesa_2017} as the reference dataset.
The flow features a Reynolds number at the Taylor microscale $\lambda$ (often used in turbulence, see~\cite{pope_2000}) of $\mathrm{Re}_{\lambda}=315$ on a periodic cubic domain of $(2\pi)^3$, that is mapped to a discretized computational domain of $(1024)^3$.
The code is available at \textcolor{violet}{\url{https://github.com/stefanopisoni/TN_Turbulence}}.

\begin{figure*}[t]
    \centering
    \includegraphics[width=\textwidth]{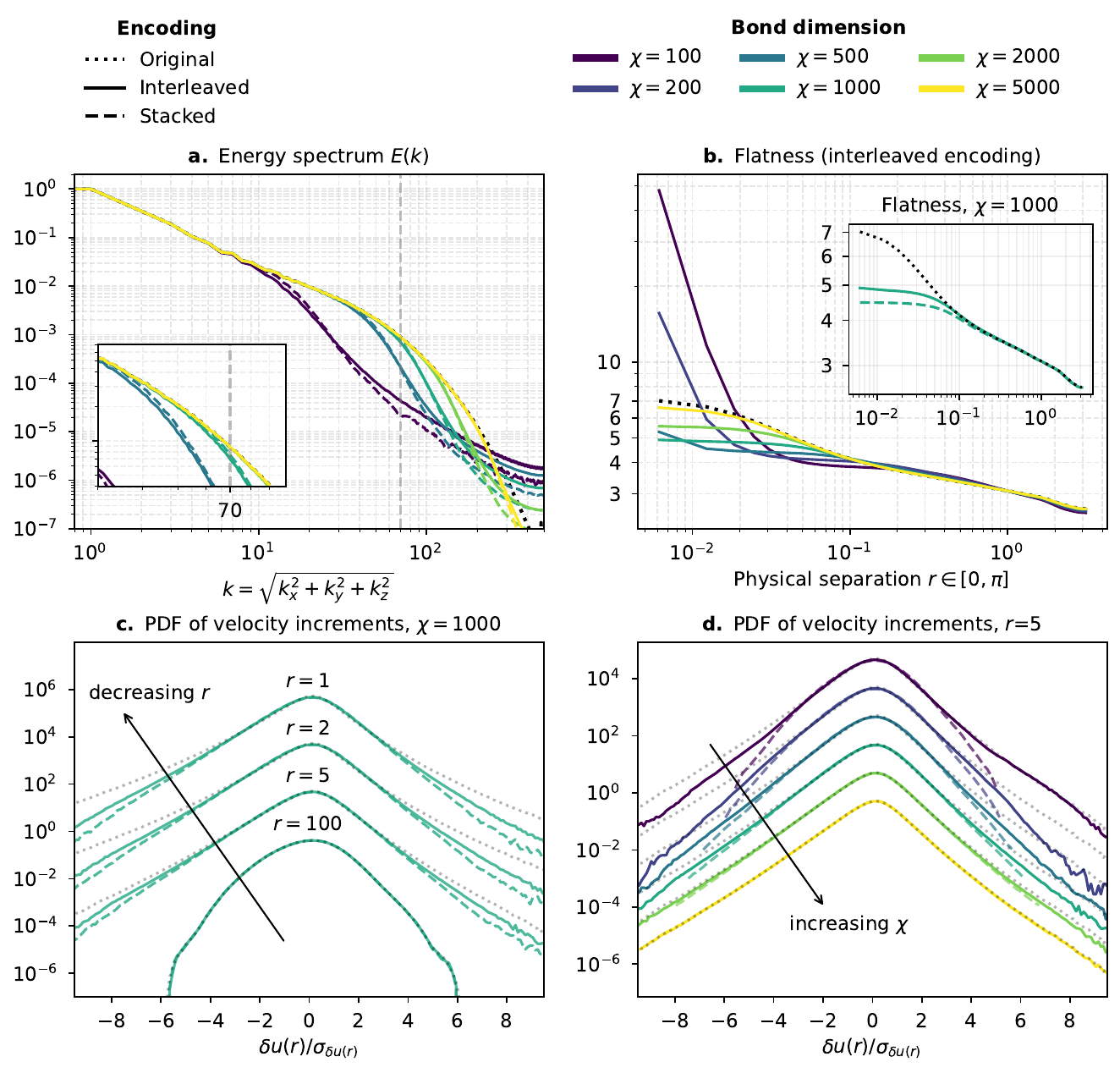}
    \caption{
    \textbf{\textit{Single snapshot compression:} Turbulent statistical metrics for the compressed TT snapshots.}
    The snapshots have been compressed and projected to satisfy the divergence-free condition. This is achieved through an iterative scheme that alternates between the compression and projection steps, for a total of 10 iterations. We end with a compression step, such that the resulting TTs have exactly the reported $\chi$.
    \textbf{a.} Kinetic turbulent energy spectrum as a function of the wave-number magnitude for various values of $\chi$ and the two possible encodings.
    The plots are on a log-log scale, and the approximately linear region corresponds to the inertial range, with a power-law decay given by Eq.~\eqref{eq:E_k}. We note that $\chi = 1000$ already reproduces entirely the inertial range for both the encodings, with the compressed spectra detaching from the original one (dotted line) around $k=70$. The inset shows a zoomed version of the main plot around that region.
    $\chi = 1000$ corresponds to a TT with only the $2.2\%$ of the total number of parameters required for its dense-vector representation. 
    \textbf{b.} The main plot reports the flatness, as per Eq.~\eqref{eq:flatness}, for the interleaved encoding and various bond dimensions. The inset shows the flatness for both the encodings at $\chi = 1000$.
    We see that for very small separation distances the curves deviate from the ground truth, either by overshooting (for low $\chi$s) or by undershooting (for large $\chi$s). We explain this behavior in the main text.
    \textbf{c.} PDF of the longitudinal velocity increments with $\chi = 1000$ for various separation distances. Here, we report the separation distances in terms of computational unit cells for clarity.
    The curves are shifted for the sake of clarity. We note that the interleaved encoding captures the non-Gaussian statistics that emerge for small separations better than the stacked one.
    \textbf{d.} PDF of the velocity increments for a given separation distance ($r=5$) and various bond dimensions. We again observe that the interleaved encoding captures the non-Gaussian statistics better than the stacked one, especially for moderate bond dimensions. We also observe, as expected, that we converge to the ground-truth statistics as $\chi$ increases.
    }
    \label{fig:single_snap_stat}
\end{figure*}

\subsection{Single snapshot compression}\label{subsection:Single snapshot compression}
In this section, we analyze the extent to which the TT representation is able to compress turbulent snapshots. To this end, we consider the following statistical metrics: 
the energy spectrum (Eq.~\eqref{eq:E_k}),
the flatness (Eq.~\eqref{eq:flatness}),
and the PDF of the longitudinal velocity increments.
For all of these metrics, we compare the TT-compressed snapshots with the original snapshots across various bond dimensions and two encodings: stacked and interleaved. In Appendix~\ref{app:Concatenated encoding}, we additionally present the energy spectrum for the concatenated encoding.

The results for the energy spectra are reported in Fig.~\ref{fig:single_snap_stat}a. As expected, we observe that the energy spectra reproduced by the truncated snapshots get closer to the ground truth as the bond dimension increases.
Specifically, at roughly $\chi=1000$, the spectrum already covers the entire flat region of the inertial range for both the encodings.
Smaller bond dimensions also reproduce good portions of the spectrum, with the high frequency part better captured by the \textit{interleaved} encoding.
We note that $\chi=1000$ retains only the $2.2\%$ of the total number of parameters, achieving a remarkable compression.
We further comment that the resolved wave number for $\chi=1000$ is around $k=70$, that is where the reconstructed spectrum detaches from the original one.

In Fig.~\ref{fig:single_snap_stat}b., we report the flatness of the truncated snapshots for the interleaved encoding and for various bond dimensions. The inset shows both the encodings at $\chi=1000$.
The figure highlights the effect of the interleaved TT compression on the intermittent nature of turbulence.
To further characterize the intermittency of the compressed snapshots for the two different encodings, we report in Fig.~\ref{fig:single_snap_stat}c. and d. the PDF of the longitudinal velocity increments.
From our analysis, we conclude that the two proposed encodings behave quite differently, with the interleaved one capturing the higher-order statistics much better than the stacked one. In fact, the stacked encoding tends to smooth out the velocity field dramatically, especially at lower values of $\chi$, while the interleaved one allows non-Gaussian statistics to occur.
However, we also note that the interleaved encoding seems to introduce artificial discontinuities or fluctuations at very small scales when a low bond dimension ($\chi < 500$) is employed, as marked by the overshooting of the two curves in Fig.~\ref{fig:single_snap_stat}b. well above the ground truth values.

This fact deserves a more detailed comment. We understand that there exists a fundamental trade-off between bond dimension truncation and divergence-free projection. In fact, a TT compression introduces discontinuities that cannot be balanced out by just enforcing a global divergence-free condition on the three TTs representing the three vector components of the velocity field.
Then, at low bond dimensions and small separation distances, this effect is stronger and causes the overshooting of the flatness observed in Fig~\ref{fig:single_snap_stat}b.

\subsection{3D TT-based solver}\label{subsection:3D MPS-based solver}

In this section, we aim at describing and validating the 3D TT-based solver, which is an extension of the 2D solver introduced in~\cite{peddinti}. The solver is based on a finite-differences approach and fundamental TT truncation and contraction schemes that feature internal routines, such as DMRG-type algorithms for solving linear systems when necessary.
After discretization, we solve equations~\eqref{eq:NS_eq_a} and~\eqref{eq:NS_eq_b} by a standard Chorin's projection scheme, where we first evolve the solution with explicit Euler time stepping and subsequently correct the solution to satisfy the divergence-free condition. This last step involves solving a Poisson-like equation for pressure, which dominates the solver's algorithmic complexity. In appendix~\ref{app:QTT-based solver}, we provide details about the 3D solver.

The simulations are performed in an empty 3D cubic domain with periodic boundary conditions (PBC), a maximum fixed bond dimension $\chi = 100$, and $N=30$. All the simulation parameters are equivalent to those of the dataset considered~\cite{Cardesa_2017}, except for the time step and the external forcing term that we do not include in Eq.~\eqref{eq:NS_eq_a}. In fact, our simulations use an explicit Euler time-stepping scheme, which is constrained by standard convergence criteria. In particular, we need to satisfy the Courant–Friedrichs–Lewy (CFL) condition, which requires that $\frac{\delta v \Delta t}{\Delta x} \leq 1$, where $\delta v$ is the maximal absolute value of velocity fluctuations. For this reason, we set the time step to be 10 times smaller than that used in the reference dataset, achieving $\mathrm{CFL} = 0.16$.

\begin{figure}
    \centering
    \includegraphics[width=\linewidth]{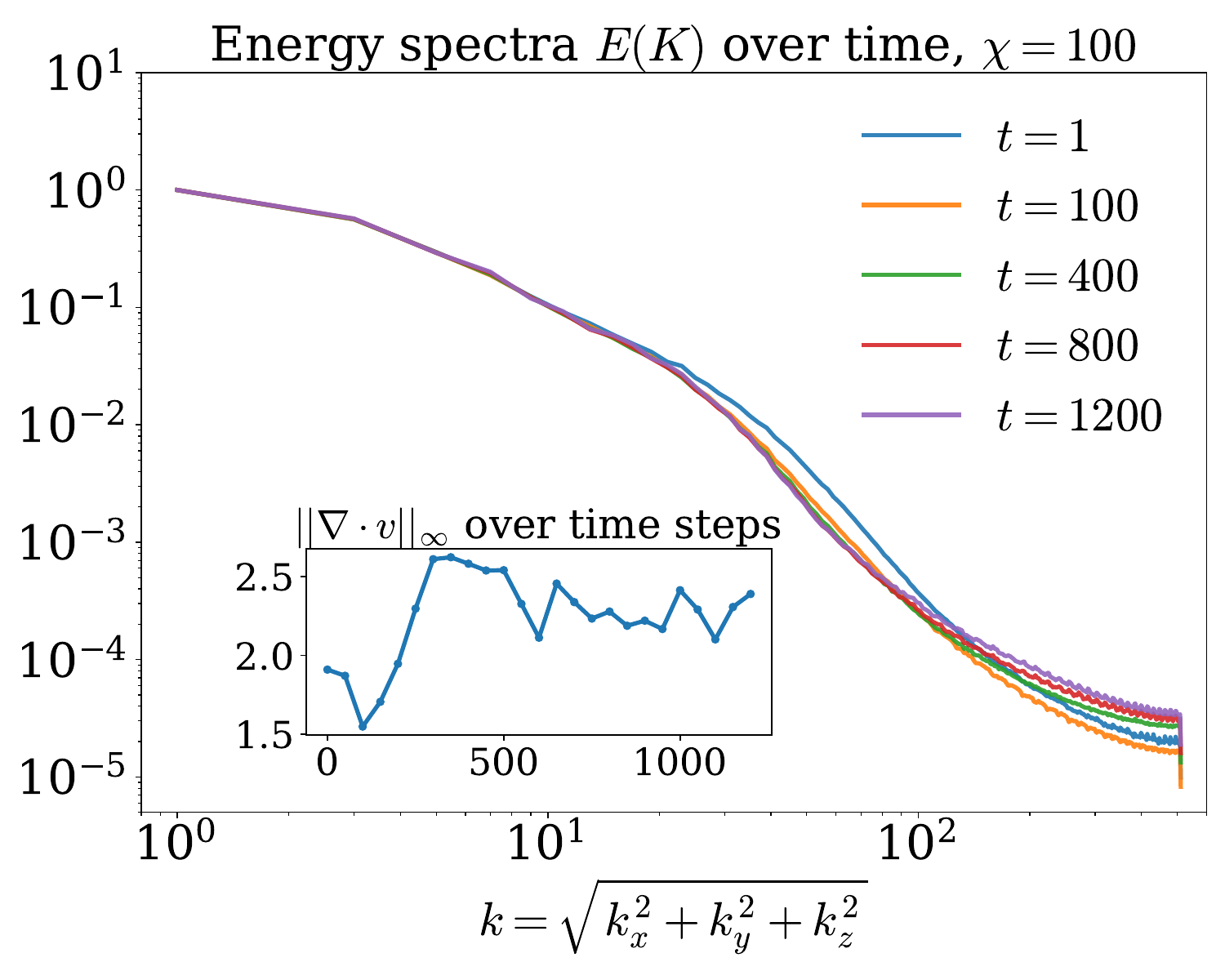}
    \caption{
    \textbf{\textit{3D TT-based solver:} $E(K)$ and divergence $L^{\infty}$-norm during the 3D TT time evolution.}
    We show the energy spectrum $E(k)$ as a function of the wave number $k$ (main figure) and the $L^{\infty}$-norm of the divergence (inset) over time.
    We performed a total of 1200 computational time steps, with the following parameters: kinematic viscosity $\nu = 0.00067$; time step $\Delta t = 0.0002$; number of tensor cores per spatial dimension $N_x = N_y = N_z = 10$; Reynolds number at the Taylor micro scale $\mathrm{Re}_{\lambda}=315$.
    These simulations use only the $0.03\%$ of the total number of parameters needed for the full vector representation.
    The number of time steps simulated is equivalent to 120 time steps in the original dataset presented in the main text~\cite{Cardesa_2017}.
    This evolution time roughly corresponds to 9-10 Kolmogorov turnover times.
    The TTs used have a low bond dimension, $\chi = 100$.
    The low bond dimension, and the absence of a forcing term in the Eqs.~\eqref{eq:NS}, explains the small deviations from the expected shape for $E(k)$.
    }
    \label{fig:3D_time_evo}
\end{figure}

\begin{figure*}
    \centering
    \includegraphics[width=\textwidth]{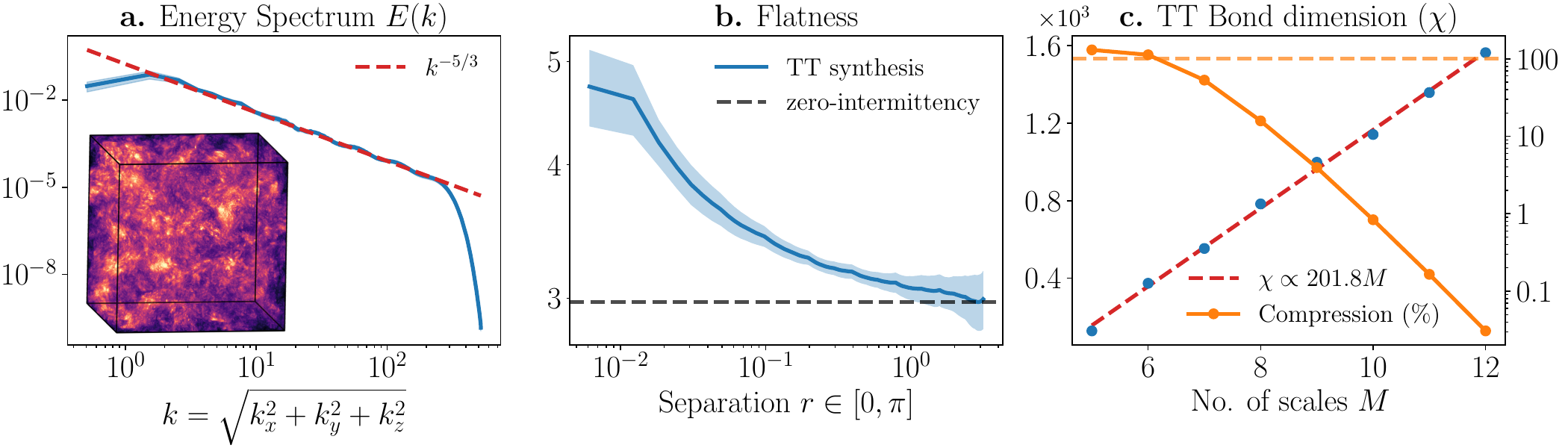}
    \caption{\textbf{\textit{TT snapshots synthesis:}
    Turbulence metrics for the synthetic generated snapshots with the proposed TT algorithm.} \textbf{a.} shows the energy spectrum $E(k)$ plotted against wavenumber $k = |\mathbf{k}|$ on a log–log scale, with the dashed red line indicating $k^{-5/3}$ power law decay. We also plot a 3D snapshot of the velocity magnitude of the synthetic turbulent field. \textbf{b.} presents the flatness (kurtosis) of velocity increments (Eq.~\eqref{eq:flatness}) as a function of physical separation $r$, on a log-log scale; The deviation from the flat line indicates the presence of intermittency, or non-Gaussianity, in the velocity fluctuations. \textbf{c.} displays the TT bond dimension $\chi$ versus the number of scales $M$ (one third of the total TT tensors), together with the dashed linear fit $\chi \propto 201.8 M$, showing a linear growth. Moreover, we show the compression rates achieved by the TT-generated snapshots with respect to the full vector representations as a function of $M$.
    We average over 20 synthetic snapshots with random seeds, and the shaded regions denote one standard deviation.}
    \label{fig:synthetic_turbulence}
\end{figure*}

To initialize the evolution, we use the compressed TT representation of a snapshot from the turbulent dataset~\cite{Cardesa_2017} as the initial condition.
During the evolution, we check whether the typical behavior of $E(k)$, expected at $\chi=100$, is well reproduced.
The incompressible flow setting allows us to further check the numerical stability of the solver by looking at the total divergence of the vector field, $\nabla \cdot \boldsymbol{v}$, over time. Therefore, we also compute the divergence and look at its $L^{\infty}$-norm as an additional sanity check.
Given that the analysis for the single snapshot compression in Sec.~\ref{subsection:Single snapshot compression} found the \textit{interleaved} encoding to be the more appropriate one, we only report the time evolution for this encoding in what follows.
The results are described and summarized in Fig.\ref{fig:3D_time_evo}.
We note the drastic compression achieved by these simulations: only the $0.03\%$ of the total number of parameters is used, compared to the reference dataset.
We observe that the energy spectrum slightly deviates over time from its expected shape at $\chi=100$. This might be due to the limited $\chi$ used for the simulations as well as the absence of a forcing term in Eqs.~\eqref{eq:NS}.
However, from a numerical point of view, the TT-solver appears to be stable and consistent, as the total divergence does not blow up over time.

We conclude by pointing out that TT-based simulations with higher $\chi$ values are needed to compete with state-of-the-art DNS. With this respect, we point out that the limited bond dimension $\chi=100$ is dictated by the memory consumption of the current implementation of the TT-based solver. Future improvements to the solver might enable larger bond dimensions, opening the way for DNS simulations at mesh sizes beyond the reach of standard CFD approaches.
Currently, another limiting factor is the small time step, which is constrained by stability criteria. This limitation can be overcome by replacing the explicit Euler scheme with an implicit method.

\subsection{TT snapshots synthesis}\label{subsection:Synthetic_turbulence}

We propose a constructive algorithm to generate 3D snapshots that exhibit turbulent-like features within the TT representation. Specifically, the snapshots satisfy the divergence‐free condition (Eq.~\eqref{eq:NS_eq_b}), the correct power law decay in the energy spectrum (Eq.~\eqref{eq:E_k}), and nonzero intermittency quantified by the flatness, Eq.~\eqref{eq:flatness}.
By constructing synthetic snapshots directly as TTs, our algorithm reduces both computational and memory costs, thereby enabling rapid generation of high-resolution flow fields for real-time applications. This is particularly useful for many applications in which the rapid generation of turbulent-like flows is demanded but often unfeasible due to fundamental limitations of DNS.
State-of-the-art methods for generating turbulent-like snapshots include Fourier spectral synthesis, wavelet decomposition, and multiscale cascade models~\cite{Kim2008-gh,Scotti1997,Scotti1999}. 
In this work, we follow a multiscale cascade approach, since its construction is well-suited for the TT representation in interleaved encoding. 
A multiscale cascade model begins by generating random fields at each subdivision level (or scale) $m$ of the domain. Then, each of these fields is interpolated up to the desired scale $M$. The final field is an additive or multiplicative combination of them, weighted by the appropriate amplitudes.
To enforce incompressibility of the synthesized velocity field $\boldsymbol{v}$, we introduce an auxiliary vector potential $\boldsymbol{A}$ called the stream function, such that $\boldsymbol{v}~=~\nabla\times\boldsymbol{A}$. Hence, we directly generate the derivatives of the stream function via:
\begin{equation} \label{eq:turb_ansatz}
    \partial_i A_j(x,y,z) = \sum_{m=2}^{M-1} \omega_m G^{m}_{i,j} (x,y,z),
\end{equation}
where the sum starts from the second scale $m=2$, i.e. $4 \times 4 \times 4$ grid, and terminates at the second-last one $M-1$. 
In Eq.~\eqref{eq:turb_ansatz}, the amplitudes $\omega_m$ are chosen to reproduce the hierarchical order of the cascade mechanism, as they control the energy injected at each scale $m$.
Specifically, according to Eq.~\eqref{eq:vel_fluc_p}, choosing $\omega_m = 2^{-m/3}$ allows us to reproduce Kolmogorov spectrum.
The $m$-th term, $G^{m}_{i,j}$, is built as shown in Fig.~\ref{fig:diagram} C: First, for each $j$, we initialize a TT with $3m$ tensors with bond dimension $\chi_G$. The entries of each tensor are sampled from a normal distribution $\mathcal{N}(0,\sigma^2)$, where $\sigma$ is chosen such that the variance of the corresponding random field with $2^{m}\times2^{m}\times2^{m}$ entries is $\sim 1$.
Second, for each $i$, we obtain a TT of the derivative $\partial_i$ by extending the number of tensors from $3m$ to $3M$. This extension to the final scale $M$ with periodic boundary conditions is done via our novel TT interpolation algorithm, as described in~\cite{siddhartha_2025}.

By construction, our new approach allows us to control the bond dimension of the synthesized field.
Specifically, each generated $G_{i,j}^m$ has a bond dimension scaling linearly with the number of tensors. Moreover, by leveraging the sub-additivity of the TT ranks, the full construction in Eq.~\eqref{eq:turb_ansatz} yields a bond dimension $\chi$ that scales linearly with the total number of TT tensors $N$.
Remarkably, this novel technique~\cite{siddhartha_2025} reduces the bond dimension of the TT representation of the velocity field by analytically constructing $\partial_i A_j$, rather than computing $\boldsymbol{A}$ and then applying an approximate derivative TT operator.

To numerically benchmark our synthetic turbulence generator algorithm, the initial noisy TTs are generated with $\chi_G=10$, and we set $M=10$.
We report the features of our synthetic field in Fig.~\ref{fig:synthetic_turbulence}, where we can observe that we are able to reproduce the correct power energy spectrum within the allowed frequency range for our finite lattice. We also compute the flatness of our synthetic field and observe intermittent behavior. 
Finally, we explicitly show that the bond dimension of the TT representation of the synthetic velocity field scales linearly with $M$, achieving an exponential memory compression.
We remark that this is not a physical turbulent snapshot, but rather a multi-scale signal that exhibits turbulent-like features.

\section{Conclusions}
\label{section:Conclusions}
We conducted a detailed investigation of TTs applied to turbulent flows. This consisted of three aspects: turbulent snapshot TT-compression, TT-based simulation of turbulent flows, and the generation of 3D TT snapshots.
The analysis of compressed turbulent snapshots is, to our knowledge, the most complete benchmark of TT encodings of fully developed turbulent flows. Retaining only $2.2\%$ of the number of parameters, TTs still successfully reproduce the key turbulence features of the original snapshots for a DNS with $Re_{\lambda} = 315$ on a 3D mesh with $2^{30}$ grid points.
Interestingly, we find that interleaved encoding outperforms stacked encoding across all turbulent metrics.
As for time evolution, our benchmark on the same turbulence dataset constitutes an unprecedented flow regime for TT methods. Our solver operates entirely within the TT representation with its memory scaling polylogarithmically with the mesh size.
Finally, our synthetic snapshot generator demonstrates that it is possible to efficiently construct a low bond-dimension TT field that reproduces some of the characteristic properties of a turbulent flow.
This algorithm features linear scaling of the bond dimension with the number of spatial scales and is equipped with a novel TT interpolation technique introduced in~\cite{siddhartha_2025}.

Although the simulated flows considered here do not have any immersed bodies, we note that our 3D solver is fully compatible with the TT mask framework presented in~\cite{peddinti}. In this regard, the TT interpolation scheme provides a new way to construct such masks in 3D, as demonstrated in~\cite{siddhartha_2025}.
Moreover, further improvements to the solver might be possible, such as migrating to implicit time-stepping schemes and optimizing the TT sub-routines. This may both speed up the simulations and allow for larger values of $\chi$. Additionally, we remark that other tensor network encodings could also be explored in the context of turbulent flows.

\bibliography{arXiv_V2}

\providecommand{\noopsort}[1]{}\providecommand{\singleLetter}[1]{#1}%
\begin{thebibliography}{37}%
\makeatletter
\providecommand \@ifxundefined [1]{%
 \@ifx{#1\undefined}
}%
\providecommand \@ifnum [1]{%
 \ifnum #1\expandafter \@firstoftwo
 \else \expandafter \@secondoftwo
 \fi
}%
\providecommand \@ifx [1]{%
 \ifx #1\expandafter \@firstoftwo
 \else \expandafter \@secondoftwo
 \fi
}%
\providecommand \natexlab [1]{#1}%
\providecommand \enquote  [1]{``#1''}%
\providecommand \bibnamefont  [1]{#1}%
\providecommand \bibfnamefont [1]{#1}%
\providecommand \citenamefont [1]{#1}%
\providecommand \href@noop [0]{\@secondoftwo}%
\providecommand \href [0]{\begingroup \@sanitize@url \@href}%
\providecommand \@href[1]{\@@startlink{#1}\@@href}%
\providecommand \@@href[1]{\endgroup#1\@@endlink}%
\providecommand \@sanitize@url [0]{\catcode `\\12\catcode `\$12\catcode `\&12\catcode `\#12\catcode `\^12\catcode `\_12\catcode `\%12\relax}%
\providecommand \@@startlink[1]{}%
\providecommand \@@endlink[0]{}%
\providecommand \url  [0]{\begingroup\@sanitize@url \@url }%
\providecommand \@url [1]{\endgroup\@href {#1}{\urlprefix }}%
\providecommand \urlprefix  [0]{URL }%
\providecommand \Eprint [0]{\href }%
\providecommand \doibase [0]{https://doi.org/}%
\providecommand \selectlanguage [0]{\@gobble}%
\providecommand \bibinfo  [0]{\@secondoftwo}%
\providecommand \bibfield  [0]{\@secondoftwo}%
\providecommand \translation [1]{[#1]}%
\providecommand \BibitemOpen [0]{}%
\providecommand \bibitemStop [0]{}%
\providecommand \bibitemNoStop [0]{.\EOS\space}%
\providecommand \EOS [0]{\spacefactor3000\relax}%
\providecommand \BibitemShut  [1]{\csname bibitem#1\endcsname}%
\let\auto@bib@innerbib\@empty
\bibitem [{\citenamefont {Peddinti}\ \emph {et~al.}(2024)\citenamefont {Peddinti}, \citenamefont {Pisoni}, \citenamefont {Marini}, \citenamefont {Lott}, \citenamefont {Argentieri}, \citenamefont {Tiunov},\ and\ \citenamefont {Aolita}}]{peddinti}%
  \BibitemOpen
  \bibfield  {author} {\bibinfo {author} {\bibfnamefont {R.~D.}\ \bibnamefont {Peddinti}}, \bibinfo {author} {\bibfnamefont {S.}~\bibnamefont {Pisoni}}, \bibinfo {author} {\bibfnamefont {A.}~\bibnamefont {Marini}}, \bibinfo {author} {\bibfnamefont {P.}~\bibnamefont {Lott}}, \bibinfo {author} {\bibfnamefont {H.}~\bibnamefont {Argentieri}}, \bibinfo {author} {\bibfnamefont {E.}~\bibnamefont {Tiunov}},\ and\ \bibinfo {author} {\bibfnamefont {L.}~\bibnamefont {Aolita}},\ }\bibfield  {title} {\bibinfo {title} {Quantum-inspired framework for computational fluid dynamics},\ }\href {https://doi.org/10.1038/s42005-024-01623-8} {\bibfield  {journal} {\bibinfo  {journal} {Communications Physics}\ }\textbf {\bibinfo {volume} {7}},\ \bibinfo {pages} {135} (\bibinfo {year} {2024})}\BibitemShut {NoStop}%
\bibitem [{\citenamefont {Pope}(2000)}]{pope_2000}%
  \BibitemOpen
  \bibfield  {author} {\bibinfo {author} {\bibfnamefont {S.~B.}\ \bibnamefont {Pope}},\ }\href {https://doi.org/10.1017/CBO9780511840531} {\emph {\bibinfo {title} {Turbulent Flows}}}\ (\bibinfo  {publisher} {Cambridge University Press},\ \bibinfo {address} {New York},\ \bibinfo {year} {2000})\BibitemShut {NoStop}%
\bibitem [{\citenamefont {Han}\ \emph {et~al.}(2018)\citenamefont {Han}, \citenamefont {Wang}, \citenamefont {Fan}, \citenamefont {Wang},\ and\ \citenamefont {Zhang}}]{han2018unsupervised}%
  \BibitemOpen
  \bibfield  {author} {\bibinfo {author} {\bibfnamefont {Z.-Y.}\ \bibnamefont {Han}}, \bibinfo {author} {\bibfnamefont {J.}~\bibnamefont {Wang}}, \bibinfo {author} {\bibfnamefont {H.}~\bibnamefont {Fan}}, \bibinfo {author} {\bibfnamefont {L.}~\bibnamefont {Wang}},\ and\ \bibinfo {author} {\bibfnamefont {P.}~\bibnamefont {Zhang}},\ }\bibfield  {title} {\bibinfo {title} {Unsupervised generative modeling using matrix product states},\ }\href@noop {} {\bibfield  {journal} {\bibinfo  {journal} {Physical Review X}\ }\textbf {\bibinfo {volume} {8}},\ \bibinfo {pages} {031012} (\bibinfo {year} {2018})}\BibitemShut {NoStop}%
\bibitem [{\citenamefont {Verstraete}\ \emph {et~al.}(2008)\citenamefont {Verstraete}, \citenamefont {Murg},\ and\ \citenamefont {Cirac}}]{verstraete2008matrix}%
  \BibitemOpen
  \bibfield  {author} {\bibinfo {author} {\bibfnamefont {F.}~\bibnamefont {Verstraete}}, \bibinfo {author} {\bibfnamefont {V.}~\bibnamefont {Murg}},\ and\ \bibinfo {author} {\bibfnamefont {J.~I.}\ \bibnamefont {Cirac}},\ }\bibfield  {title} {\bibinfo {title} {Matrix product states, projected entangled pair states, and variational renormalization group methods for quantum spin systems},\ }\href@noop {} {\bibfield  {journal} {\bibinfo  {journal} {Advances in Physics}\ }\textbf {\bibinfo {volume} {57}},\ \bibinfo {pages} {143} (\bibinfo {year} {2008})}\BibitemShut {NoStop}%
\bibitem [{\citenamefont {Or{\'u}s}(2014)}]{orus2014practical}%
  \BibitemOpen
  \bibfield  {author} {\bibinfo {author} {\bibfnamefont {R.}~\bibnamefont {Or{\'u}s}},\ }\bibfield  {title} {\bibinfo {title} {A practical introduction to tensor networks: matrix product states and projected entangled pair states},\ }\href@noop {} {\bibfield  {journal} {\bibinfo  {journal} {Annals of Physics}\ }\textbf {\bibinfo {volume} {349}},\ \bibinfo {pages} {117} (\bibinfo {year} {2014})}\BibitemShut {NoStop}%
\bibitem [{\citenamefont {Oseledets}(2011)}]{oseledets2011tensortrain}%
  \BibitemOpen
  \bibfield  {author} {\bibinfo {author} {\bibfnamefont {I.~V.}\ \bibnamefont {Oseledets}},\ }\bibfield  {title} {\bibinfo {title} {Tensor-train decomposition},\ }\href@noop {} {\bibfield  {journal} {\bibinfo  {journal} {SIAM Journal on Scientific Computing}\ }\textbf {\bibinfo {volume} {33}},\ \bibinfo {pages} {2295} (\bibinfo {year} {2011})}\BibitemShut {NoStop}%
\bibitem [{\citenamefont {Khoromskij}(2011)}]{Khoromskij_2011}%
  \BibitemOpen
  \bibfield  {author} {\bibinfo {author} {\bibfnamefont {B.~N.}\ \bibnamefont {Khoromskij}},\ }\bibfield  {title} {\bibinfo {title} {$\mathcal{O}(d\log{N})$-quantics approximation of {N}-d tensors in high-dimensional numerical modeling},\ }\href@noop {} {\bibfield  {journal} {\bibinfo  {journal} {Constructive Approximation}\ }\textbf {\bibinfo {volume} {34}},\ \bibinfo {pages} {257–280} (\bibinfo {year} {2011})}\BibitemShut {NoStop}%
\bibitem [{\citenamefont {White}(1992)}]{White_PRL_1992}%
  \BibitemOpen
  \bibfield  {author} {\bibinfo {author} {\bibfnamefont {S.~R.}\ \bibnamefont {White}},\ }\bibfield  {title} {\bibinfo {title} {Density matrix formulation for quantum renormalization groups},\ }\href {https://doi.org/10.1103/PhysRevLett.69.2863} {\bibfield  {journal} {\bibinfo  {journal} {Physical Review Letters}\ }\textbf {\bibinfo {volume} {69}},\ \bibinfo {pages} {2863} (\bibinfo {year} {1992})}\BibitemShut {NoStop}%
\bibitem [{\citenamefont {Latorre}(2005)}]{Image_Compression}%
  \BibitemOpen
  \bibfield  {author} {\bibinfo {author} {\bibfnamefont {J.}~\bibnamefont {Latorre}},\ }\bibfield  {title} {\bibinfo {title} {Image compression and entanglement},\ }\href@noop {} {\bibfield  {journal} {\bibinfo  {journal} {abs/quant-ph/0510031}\ } (\bibinfo {year} {2005})}\BibitemShut {NoStop}%
\bibitem [{\citenamefont {Kastoryano}\ and\ \citenamefont {Pancotti}(2022)}]{option_pricing}%
  \BibitemOpen
  \bibfield  {author} {\bibinfo {author} {\bibfnamefont {M.}~\bibnamefont {Kastoryano}}\ and\ \bibinfo {author} {\bibfnamefont {N.}~\bibnamefont {Pancotti}},\ }\bibfield  {title} {\bibinfo {title} {A highly efficient tensor network algorithm for multi-asset fourier options pricing},\ }\href@noop {} {\bibfield  {journal} {\bibinfo  {journal} {arXiv preprint arXiv:2203.02804}\ } (\bibinfo {year} {2022})}\BibitemShut {NoStop}%
\bibitem [{\citenamefont {Gourianov}\ \emph {et~al.}(2022)\citenamefont {Gourianov}, \citenamefont {Lubasch}, \citenamefont {Dolgov}, \citenamefont {van~den Berg}, \citenamefont {Babaee}, \citenamefont {Givi}, \citenamefont {Kiffner},\ and\ \citenamefont {Jaksch}}]{gourianov2022quantum}%
  \BibitemOpen
  \bibfield  {author} {\bibinfo {author} {\bibfnamefont {N.}~\bibnamefont {Gourianov}}, \bibinfo {author} {\bibfnamefont {M.}~\bibnamefont {Lubasch}}, \bibinfo {author} {\bibfnamefont {S.}~\bibnamefont {Dolgov}}, \bibinfo {author} {\bibfnamefont {Q.~Y.}\ \bibnamefont {van~den Berg}}, \bibinfo {author} {\bibfnamefont {H.}~\bibnamefont {Babaee}}, \bibinfo {author} {\bibfnamefont {P.}~\bibnamefont {Givi}}, \bibinfo {author} {\bibfnamefont {M.}~\bibnamefont {Kiffner}},\ and\ \bibinfo {author} {\bibfnamefont {D.}~\bibnamefont {Jaksch}},\ }\bibfield  {title} {\bibinfo {title} {A quantum-inspired approach to exploit turbulence structures},\ }\href@noop {} {\bibfield  {journal} {\bibinfo  {journal} {Nature Computational Science}\ }\textbf {\bibinfo {volume} {2}},\ \bibinfo {pages} {30} (\bibinfo {year} {2022})}\BibitemShut {NoStop}%
\bibitem [{\citenamefont {Peddinti}\ \emph {et~al.}(2025)\citenamefont {Peddinti}, \citenamefont {Pisoni}, \citenamefont {Tiunov}, \citenamefont {Marini},\ and\ \citenamefont {Aolita}}]{peddinti2025technicalreportquantuminspiredsolver}%
  \BibitemOpen
  \bibfield  {author} {\bibinfo {author} {\bibfnamefont {R.~D.}\ \bibnamefont {Peddinti}}, \bibinfo {author} {\bibfnamefont {S.}~\bibnamefont {Pisoni}}, \bibinfo {author} {\bibfnamefont {E.}~\bibnamefont {Tiunov}}, \bibinfo {author} {\bibfnamefont {A.}~\bibnamefont {Marini}},\ and\ \bibinfo {author} {\bibfnamefont {L.}~\bibnamefont {Aolita}},\ }\href {https://arxiv.org/abs/2506.03833} {\bibinfo {title} {Technical report on a quantum-inspired solver for simulating compressible flows}} (\bibinfo {year} {2025}),\ \Eprint {https://arxiv.org/abs/2506.03833} {arXiv:2506.03833 [physics.flu-dyn]} \BibitemShut {NoStop}%
\bibitem [{\citenamefont {Kiffner}\ and\ \citenamefont {Jaksch}(2023)}]{kiffner_jaksch2023tensor}%
  \BibitemOpen
  \bibfield  {author} {\bibinfo {author} {\bibfnamefont {M.}~\bibnamefont {Kiffner}}\ and\ \bibinfo {author} {\bibfnamefont {D.}~\bibnamefont {Jaksch}},\ }\bibfield  {title} {\bibinfo {title} {Tensor network reduced order models for wall-bounded flows},\ }\Eprint {https://arxiv.org/abs/2303.03010} {arXiv:2303.03010 [Physics.flu-dyn]}  (\bibinfo {year} {2023}),\ \bibinfo {note} {-}\BibitemShut {NoStop}%
\bibitem [{\citenamefont {Kornev}\ \emph {et~al.}(2023)\citenamefont {Kornev}, \citenamefont {Dolgov}, \citenamefont {Pinto}, \citenamefont {Pflitsch}, \citenamefont {Perelshtein},\ and\ \citenamefont {Melnikov}}]{kornev2023chemicalmixer}%
  \BibitemOpen
  \bibfield  {author} {\bibinfo {author} {\bibfnamefont {E.}~\bibnamefont {Kornev}}, \bibinfo {author} {\bibfnamefont {S.}~\bibnamefont {Dolgov}}, \bibinfo {author} {\bibfnamefont {K.}~\bibnamefont {Pinto}}, \bibinfo {author} {\bibfnamefont {M.}~\bibnamefont {Pflitsch}}, \bibinfo {author} {\bibfnamefont {M.}~\bibnamefont {Perelshtein}},\ and\ \bibinfo {author} {\bibfnamefont {A.}~\bibnamefont {Melnikov}},\ }\bibfield  {title} {\bibinfo {title} {Numerical solution of the incompressible {Navier-Stokes} equations for chemical mixers via quantum-inspired {Tensor Train Finite Element Method}},\ }\href@noop {} {\bibfield  {journal} {\bibinfo  {journal} {arXiv preprint arXiv:2305.10784}\ } (\bibinfo {year} {2023})}\BibitemShut {NoStop}%
\bibitem [{\citenamefont {Arenstein}\ \emph {et~al.}(2025)\citenamefont {Arenstein}, \citenamefont {Mikkelsen},\ and\ \citenamefont {Kastoryano}}]{arenstein2025fast}%
  \BibitemOpen
  \bibfield  {author} {\bibinfo {author} {\bibfnamefont {L.}~\bibnamefont {Arenstein}}, \bibinfo {author} {\bibfnamefont {M.}~\bibnamefont {Mikkelsen}},\ and\ \bibinfo {author} {\bibfnamefont {M.}~\bibnamefont {Kastoryano}},\ }\bibfield  {title} {\bibinfo {title} {Fast and flexible quantum-inspired differential equation solvers with data integration},\ }\href@noop {} {\bibfield  {journal} {\bibinfo  {journal} {arXiv preprint arXiv:2505.17046}\ } (\bibinfo {year} {2025})}\BibitemShut {NoStop}%
\bibitem [{\citenamefont {Gourianov}\ \emph {et~al.}(2025)\citenamefont {Gourianov}, \citenamefont {Givi}, \citenamefont {Jaksch},\ and\ \citenamefont {Pope}}]{Turbulent_pdf_tensor_networks_Gourianov}%
  \BibitemOpen
  \bibfield  {author} {\bibinfo {author} {\bibfnamefont {N.}~\bibnamefont {Gourianov}}, \bibinfo {author} {\bibfnamefont {P.}~\bibnamefont {Givi}}, \bibinfo {author} {\bibfnamefont {D.}~\bibnamefont {Jaksch}},\ and\ \bibinfo {author} {\bibfnamefont {S.~B.}\ \bibnamefont {Pope}},\ }\bibfield  {title} {\bibinfo {title} {Tensor networks enable the calculation of turbulence probability distributions},\ }\href {https://doi.org/10.1126/sciadv.ads5990} {\bibfield  {journal} {\bibinfo  {journal} {Science Advances}\ }\textbf {\bibinfo {volume} {11}},\ \bibinfo {pages} {eads5990} (\bibinfo {year} {2025})},\ \Eprint {https://arxiv.org/abs/https://www.science.org/doi/pdf/10.1126/sciadv.ads5990} {https://www.science.org/doi/pdf/10.1126/sciadv.ads5990} \BibitemShut {NoStop}%
\bibitem [{\citenamefont {van Hülst}\ \emph {et~al.}(2025)\citenamefont {van Hülst}, \citenamefont {Siegl}, \citenamefont {Over}, \citenamefont {Bengoechea}, \citenamefont {Hashizume}, \citenamefont {Cecile}, \citenamefont {Rung},\ and\ \citenamefont {Jaksch}}]{quantuminspiredtensornetworkfractionalstepmethod}%
  \BibitemOpen
  \bibfield  {author} {\bibinfo {author} {\bibfnamefont {N.-L.}\ \bibnamefont {van Hülst}}, \bibinfo {author} {\bibfnamefont {P.}~\bibnamefont {Siegl}}, \bibinfo {author} {\bibfnamefont {P.}~\bibnamefont {Over}}, \bibinfo {author} {\bibfnamefont {S.}~\bibnamefont {Bengoechea}}, \bibinfo {author} {\bibfnamefont {T.}~\bibnamefont {Hashizume}}, \bibinfo {author} {\bibfnamefont {M.~G.}\ \bibnamefont {Cecile}}, \bibinfo {author} {\bibfnamefont {T.}~\bibnamefont {Rung}},\ and\ \bibinfo {author} {\bibfnamefont {D.}~\bibnamefont {Jaksch}},\ }\href {https://arxiv.org/abs/2507.05222} {\bibinfo {title} {Quantum-inspired tensor-network fractional-step method for incompressible flow in curvilinear coordinates}} (\bibinfo {year} {2025}),\ \Eprint {https://arxiv.org/abs/2507.05222} {arXiv:2507.05222 [physics.flu-dyn]} \BibitemShut {NoStop}%
\bibitem [{\citenamefont {Kolmogorov}(1941)}]{kolmogorov1941local}%
  \BibitemOpen
  \bibfield  {author} {\bibinfo {author} {\bibfnamefont {A.~N.}\ \bibnamefont {Kolmogorov}},\ }\bibfield  {title} {\bibinfo {title} {The local structure of turbulence in incompressible viscous fluid for very large {Reynolds} number},\ }in\ \href@noop {} {\emph {\bibinfo {booktitle} {Dokl. Akad. Nauk. SSSR}}},\ Vol.~\bibinfo {volume} {30}\ (\bibinfo {year} {1941})\ pp.\ \bibinfo {pages} {301--303}\BibitemShut {NoStop}%
\bibitem [{\citenamefont {Cardesa}\ \emph {et~al.}(2017)\citenamefont {Cardesa}, \citenamefont {Vela-Martín},\ and\ \citenamefont {Jiménez}}]{Cardesa_2017}%
  \BibitemOpen
  \bibfield  {author} {\bibinfo {author} {\bibfnamefont {J.~I.}\ \bibnamefont {Cardesa}}, \bibinfo {author} {\bibfnamefont {A.}~\bibnamefont {Vela-Martín}},\ and\ \bibinfo {author} {\bibfnamefont {J.}~\bibnamefont {Jiménez}},\ }\bibfield  {title} {\bibinfo {title} {The turbulent cascade in five dimensions},\ }\href {https://doi.org/10.1126/science.aan7933} {\bibfield  {journal} {\bibinfo  {journal} {Science}\ }\textbf {\bibinfo {volume} {357}},\ \bibinfo {pages} {782–784} (\bibinfo {year} {2017})}\BibitemShut {NoStop}%
\bibitem [{\citenamefont {Guzman}\ \emph {et~al.}(2026)\citenamefont {Guzman}, \citenamefont {Tiunov},\ and\ \citenamefont {Aolita}}]{siddhartha_2025}%
  \BibitemOpen
  \bibfield  {author} {\bibinfo {author} {\bibfnamefont {S.~E.}\ \bibnamefont {Guzman}}, \bibinfo {author} {\bibfnamefont {E.}~\bibnamefont {Tiunov}},\ and\ \bibinfo {author} {\bibfnamefont {L.}~\bibnamefont {Aolita}},\ }\bibfield  {title} {\bibinfo {title} {Local interpolation via low-rank tensor trains},\ }\href@noop {} {\bibfield  {journal} {\bibinfo  {journal} {arXiv preprint arXiv:2601.03885}\ } (\bibinfo {year} {2026})}\BibitemShut {NoStop}%
\bibitem [{\citenamefont {Schollw{\"o}ck}(2011)}]{schollwock2011density}%
  \BibitemOpen
  \bibfield  {author} {\bibinfo {author} {\bibfnamefont {U.}~\bibnamefont {Schollw{\"o}ck}},\ }\bibfield  {title} {\bibinfo {title} {The density-matrix renormalization group in the age of matrix product states},\ }\href@noop {} {\bibfield  {journal} {\bibinfo  {journal} {Annals of Physics}\ }\textbf {\bibinfo {volume} {326}},\ \bibinfo {pages} {96} (\bibinfo {year} {2011})}\BibitemShut {NoStop}%
\bibitem [{\citenamefont {Benzi}\ and\ \citenamefont {Biferale}(2015)}]{benzi_homogeneous_2015}%
  \BibitemOpen
  \bibfield  {author} {\bibinfo {author} {\bibfnamefont {R.}~\bibnamefont {Benzi}}\ and\ \bibinfo {author} {\bibfnamefont {L.}~\bibnamefont {Biferale}},\ }\bibfield  {title} {\bibinfo {title} {Homogeneous and {Isotropic} {Turbulence}: {A} {Short} {Survey} on {Recent} {Developments}},\ }\href {https://doi.org/10.1007/s10955-015-1323-9} {\bibfield  {journal} {\bibinfo  {journal} {Journal of Statistical Physics}\ }\textbf {\bibinfo {volume} {161}},\ \bibinfo {pages} {1351} (\bibinfo {year} {2015})}\BibitemShut {NoStop}%
\bibitem [{\citenamefont {Monin}\ and\ \citenamefont {Yaglom}(2007)}]{monin2007}%
  \BibitemOpen
  \bibfield  {author} {\bibinfo {author} {\bibfnamefont {A.~S.}\ \bibnamefont {Monin}}\ and\ \bibinfo {author} {\bibfnamefont {A.~M.}\ \bibnamefont {Yaglom}},\ }\href@noop {} {\emph {\bibinfo {title} {Statistical fluid mechanics, volume I}}}\ (\bibinfo  {publisher} {Courier Corporation},\ \bibinfo {address} {New York},\ \bibinfo {year} {2007})\BibitemShut {NoStop}%
\bibitem [{\citenamefont {Monin}\ and\ \citenamefont {Yaglom}(2013)}]{monin2013statistical}%
  \BibitemOpen
  \bibfield  {author} {\bibinfo {author} {\bibfnamefont {A.~S.}\ \bibnamefont {Monin}}\ and\ \bibinfo {author} {\bibfnamefont {A.~M.}\ \bibnamefont {Yaglom}},\ }\href@noop {} {\emph {\bibinfo {title} {Statistical fluid mechanics, volume II: mechanics of turbulence}}}\ (\bibinfo  {publisher} {Courier Corporation},\ \bibinfo {address} {New York},\ \bibinfo {year} {2013})\BibitemShut {NoStop}%
\bibitem [{\citenamefont {Eyink}(2005)}]{EYINK_locality_turbulent_cascade}%
  \BibitemOpen
  \bibfield  {author} {\bibinfo {author} {\bibfnamefont {G.~L.}\ \bibnamefont {Eyink}},\ }\bibfield  {title} {\bibinfo {title} {Locality of turbulent cascades},\ }\href {https://doi.org/https://doi.org/10.1016/j.physd.2005.05.018} {\bibfield  {journal} {\bibinfo  {journal} {Physica D: Nonlinear Phenomena}\ }\textbf {\bibinfo {volume} {207}},\ \bibinfo {pages} {91} (\bibinfo {year} {2005})}\BibitemShut {NoStop}%
\bibitem [{\citenamefont {Kraichnan}(1959)}]{Kraichnan_1959_HIT_high_Reynolds}%
  \BibitemOpen
  \bibfield  {author} {\bibinfo {author} {\bibfnamefont {R.~H.}\ \bibnamefont {Kraichnan}},\ }\bibfield  {title} {\bibinfo {title} {The structure of isotropic turbulence at very high reynolds numbers},\ }\href {https://doi.org/10.1017/S0022112059000362} {\bibfield  {journal} {\bibinfo  {journal} {Journal of Fluid Mechanics}\ }\textbf {\bibinfo {volume} {5}},\ \bibinfo {pages} {497–543} (\bibinfo {year} {1959})}\BibitemShut {NoStop}%
\bibitem [{\citenamefont {Biferale}(2003)}]{biferale2003shell}%
  \BibitemOpen
  \bibfield  {author} {\bibinfo {author} {\bibfnamefont {L.}~\bibnamefont {Biferale}},\ }\bibfield  {title} {\bibinfo {title} {Shell models of energy cascade in turbulence},\ }\href@noop {} {\bibfield  {journal} {\bibinfo  {journal} {Annual review of fluid mechanics}\ }\textbf {\bibinfo {volume} {35}},\ \bibinfo {pages} {441} (\bibinfo {year} {2003})}\BibitemShut {NoStop}%
\bibitem [{\citenamefont {Arn{\'e}odo}\ \emph {et~al.}(2008)\citenamefont {Arn{\'e}odo}, \citenamefont {Benzi}, \citenamefont {Berg}, \citenamefont {Biferale}, \citenamefont {Bodenschatz}, \citenamefont {Busse}, \citenamefont {Calzavarini}, \citenamefont {Castaing}, \citenamefont {Cencini}, \citenamefont {Chevillard} \emph {et~al.}}]{universal_stats_turb_experiments_2008}%
  \BibitemOpen
  \bibfield  {author} {\bibinfo {author} {\bibfnamefont {A.}~\bibnamefont {Arn{\'e}odo}}, \bibinfo {author} {\bibfnamefont {R.}~\bibnamefont {Benzi}}, \bibinfo {author} {\bibfnamefont {J.}~\bibnamefont {Berg}}, \bibinfo {author} {\bibfnamefont {L.}~\bibnamefont {Biferale}}, \bibinfo {author} {\bibfnamefont {E.}~\bibnamefont {Bodenschatz}}, \bibinfo {author} {\bibfnamefont {A.}~\bibnamefont {Busse}}, \bibinfo {author} {\bibfnamefont {E.}~\bibnamefont {Calzavarini}}, \bibinfo {author} {\bibfnamefont {B.}~\bibnamefont {Castaing}}, \bibinfo {author} {\bibfnamefont {M.}~\bibnamefont {Cencini}}, \bibinfo {author} {\bibfnamefont {L.}~\bibnamefont {Chevillard}}, \emph {et~al.},\ }\bibfield  {title} {\bibinfo {title} {Universal intermittent properties of particle trajectories in highly turbulent flows},\ }\href@noop {} {\bibfield  {journal} {\bibinfo  {journal} {Physical review letters}\ }\textbf {\bibinfo {volume} {100}},\ \bibinfo {pages} {254504} (\bibinfo {year} {2008})}\BibitemShut {NoStop}%
\bibitem [{\citenamefont {Kim}\ \emph {et~al.}(2008)\citenamefont {Kim}, \citenamefont {Th{\"u}rey}, \citenamefont {James},\ and\ \citenamefont {Gross}}]{Kim2008-gh}%
  \BibitemOpen
  \bibfield  {author} {\bibinfo {author} {\bibfnamefont {T.}~\bibnamefont {Kim}}, \bibinfo {author} {\bibfnamefont {N.}~\bibnamefont {Th{\"u}rey}}, \bibinfo {author} {\bibfnamefont {D.}~\bibnamefont {James}},\ and\ \bibinfo {author} {\bibfnamefont {M.}~\bibnamefont {Gross}},\ }\bibfield  {title} {\bibinfo {title} {Wavelet turbulence for fluid simulation},\ }\href@noop {} {\bibfield  {journal} {\bibinfo  {journal} {ACM Trans. Graph.}\ }\textbf {\bibinfo {volume} {27}},\ \bibinfo {pages} {1} (\bibinfo {year} {2008})}\BibitemShut {NoStop}%
\bibitem [{\citenamefont {Scotti}\ and\ \citenamefont {Meneveau}(1997)}]{Scotti1997}%
  \BibitemOpen
  \bibfield  {author} {\bibinfo {author} {\bibfnamefont {A.}~\bibnamefont {Scotti}}\ and\ \bibinfo {author} {\bibfnamefont {C.}~\bibnamefont {Meneveau}},\ }\bibfield  {title} {\bibinfo {title} {Fractal model for coarse-grained nonlinear partial differential equations},\ }\href {https://doi.org/10.1103/physrevlett.78.867} {\bibfield  {journal} {\bibinfo  {journal} {Physical Review Letters}\ }\textbf {\bibinfo {volume} {78}},\ \bibinfo {pages} {867–870} (\bibinfo {year} {1997})}\BibitemShut {NoStop}%
\bibitem [{\citenamefont {Scotti}\ and\ \citenamefont {Meneveau}(1999)}]{Scotti1999}%
  \BibitemOpen
  \bibfield  {author} {\bibinfo {author} {\bibfnamefont {A.}~\bibnamefont {Scotti}}\ and\ \bibinfo {author} {\bibfnamefont {C.}~\bibnamefont {Meneveau}},\ }\bibfield  {title} {\bibinfo {title} {A fractal model for large eddy simulation of turbulent flow},\ }\href {https://doi.org/10.1016/s0167-2789(98)00266-8} {\bibfield  {journal} {\bibinfo  {journal} {Physica D: Nonlinear Phenomena}\ }\textbf {\bibinfo {volume} {127}},\ \bibinfo {pages} {198–232} (\bibinfo {year} {1999})}\BibitemShut {NoStop}%
\bibitem [{\citenamefont {Kazeev}\ and\ \citenamefont {Khoromskij}(2012)}]{kazeev2012low}%
  \BibitemOpen
  \bibfield  {author} {\bibinfo {author} {\bibfnamefont {V.~A.}\ \bibnamefont {Kazeev}}\ and\ \bibinfo {author} {\bibfnamefont {B.~N.}\ \bibnamefont {Khoromskij}},\ }\bibfield  {title} {\bibinfo {title} {Low-rank explicit {QTT} representation of the laplace operator and its inverse},\ }\href@noop {} {\bibfield  {journal} {\bibinfo  {journal} {SIAM Journal on Matrix Analysis and Applications}\ }\textbf {\bibinfo {volume} {33}},\ \bibinfo {pages} {742} (\bibinfo {year} {2012})}\BibitemShut {NoStop}%
\bibitem [{\citenamefont {Chorin}(1967)}]{chorin1967numerical}%
  \BibitemOpen
  \bibfield  {author} {\bibinfo {author} {\bibfnamefont {A.~J.}\ \bibnamefont {Chorin}},\ }\bibfield  {title} {\bibinfo {title} {The numerical solution of the {Navier-Stokes} equations for an incompressible fluid},\ }\href@noop {} {\bibfield  {journal} {\bibinfo  {journal} {Bulletin of the American Mathematical Society}\ }\textbf {\bibinfo {volume} {73}},\ \bibinfo {pages} {928} (\bibinfo {year} {1967})}\BibitemShut {NoStop}%
\bibitem [{\citenamefont {Chorin}(1968)}]{chorin1968numerical}%
  \BibitemOpen
  \bibfield  {author} {\bibinfo {author} {\bibfnamefont {A.~J.}\ \bibnamefont {Chorin}},\ }\bibfield  {title} {\bibinfo {title} {Numerical solution of the {Navier-Stokes} equations},\ }\href {https://doi.org/10.1090/s0025-5718-1968-0242392-2} {\bibfield  {journal} {\bibinfo  {journal} {Mathematics of Computation}\ }\textbf {\bibinfo {volume} {22}},\ \bibinfo {pages} {745–762} (\bibinfo {year} {1968})}\BibitemShut {NoStop}%
\bibitem [{\citenamefont {Oseledets}\ and\ \citenamefont {Dolgov}(2012)}]{oseledets2012solution}%
  \BibitemOpen
  \bibfield  {author} {\bibinfo {author} {\bibfnamefont {I.~V.}\ \bibnamefont {Oseledets}}\ and\ \bibinfo {author} {\bibfnamefont {S.~V.}\ \bibnamefont {Dolgov}},\ }\bibfield  {title} {\bibinfo {title} {Solution of linear systems and matrix inversion in the {TT}-format},\ }\href@noop {} {\bibfield  {journal} {\bibinfo  {journal} {SIAM Journal on Scientific Computing}\ }\textbf {\bibinfo {volume} {34}},\ \bibinfo {pages} {A2718} (\bibinfo {year} {2012})}\BibitemShut {NoStop}%
\bibitem [{\citenamefont {Dolgov}\ and\ \citenamefont {Savostyanov}(2014)}]{Dolgov_Savostyanov_2014}%
  \BibitemOpen
  \bibfield  {author} {\bibinfo {author} {\bibfnamefont {S.~V.}\ \bibnamefont {Dolgov}}\ and\ \bibinfo {author} {\bibfnamefont {D.~V.}\ \bibnamefont {Savostyanov}},\ }\bibfield  {title} {\bibinfo {title} {Alternating minimal energy methods for linear systems in higher dimensions},\ }\href {https://doi.org/10.1137/140953289} {\bibfield  {journal} {\bibinfo  {journal} {SIAM Journal on Scientific Computing}\ }\textbf {\bibinfo {volume} {36}},\ \bibinfo {pages} {A2248–A2271} (\bibinfo {year} {2014})}\BibitemShut {NoStop}%
\bibitem [{\citenamefont {Courant}\ \emph {et~al.}(1928)\citenamefont {Courant}, \citenamefont {Friedrichs},\ and\ \citenamefont {Lewy}}]{cfl_1928}%
  \BibitemOpen
  \bibfield  {author} {\bibinfo {author} {\bibfnamefont {R.}~\bibnamefont {Courant}}, \bibinfo {author} {\bibfnamefont {K.}~\bibnamefont {Friedrichs}},\ and\ \bibinfo {author} {\bibfnamefont {H.}~\bibnamefont {Lewy}},\ }\bibfield  {title} {\bibinfo {title} {Uber die partiellen differenzengleichungen der mathematischen physik},\ }\href {http://dx.doi.org/10.1007/BF01448839} {\bibfield  {journal} {\bibinfo  {journal} {Mathematische Annalen}\ }\textbf {\bibinfo {volume} {100}} (\bibinfo {year} {1928})}\BibitemShut {NoStop}%
\end{thebibliography}%

\clearpage

\setcounter{figure}{0}
\renewcommand\thefigure{A\arabic{figure}}

\onecolumngrid
\begin{center}
    \large \textbf{Supplementary Information}
\end{center}
\twocolumngrid

\appendix

\section{Concatenated encoding} \label{app:Concatenated encoding}

In this appendix we present and analyze the idea of encoding the vector field components $(u, v, w)$ into a single TT, stitching them together with an additional extra tensor. The physical leg of the extra tensor will then be three-dimensional to label the three different components of $\boldsymbol{v}$.
In analogy to the snapshots compression analysis, we fix a maximal bond dimension $\chi$ and truncate the augmented TT according to it.

\begin{figure}[h]
    \centering
    \includegraphics[width=\linewidth]{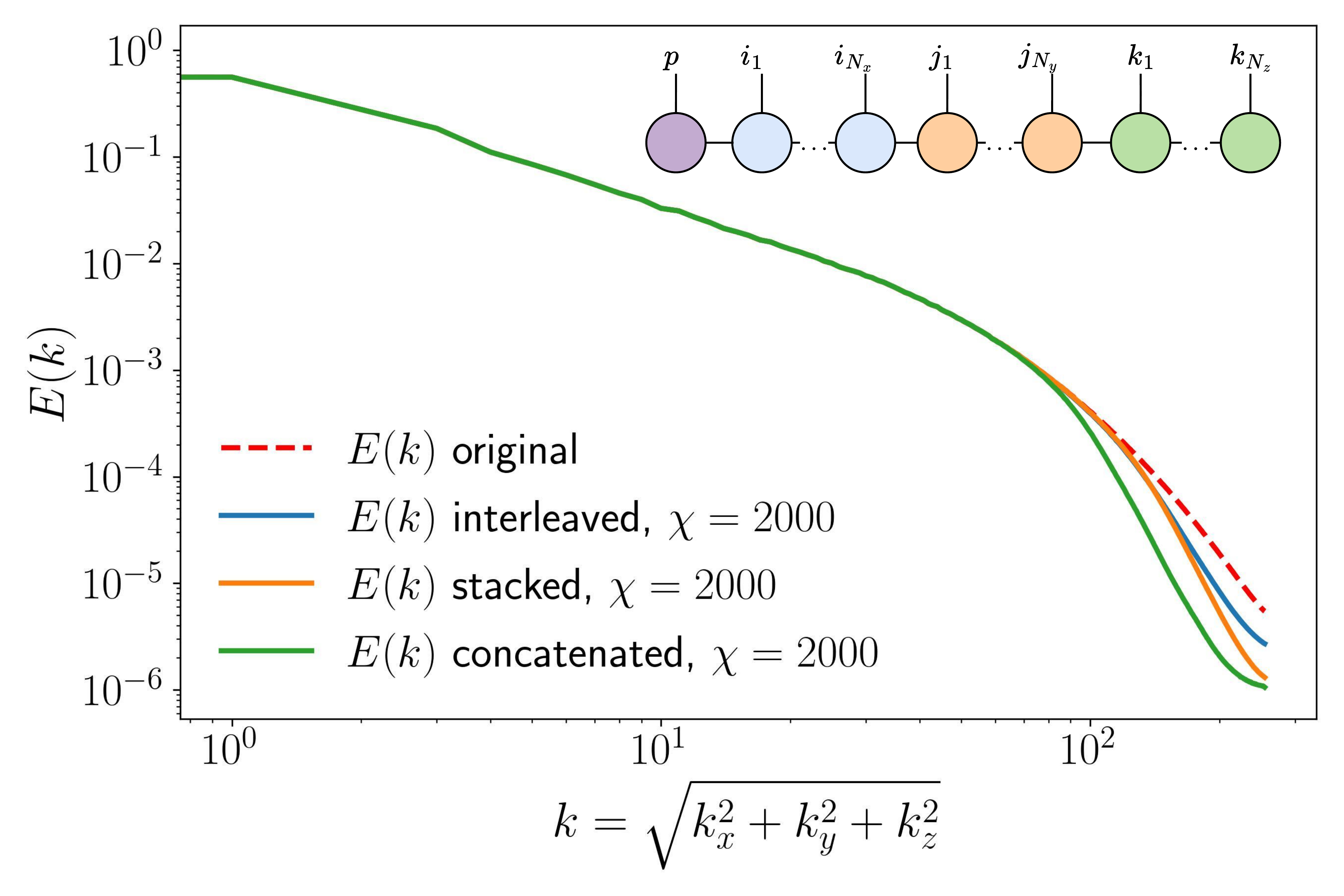}
    \caption{
    \textbf{Concatenated tensor train energy spectrum.} We plot the energy spectrum for the three different encodings introduced in Fig.~\ref{fig:encodings} for a fixed bond dimension $\chi = 2000$.
    Note that the concatenated encoding is already encoding the three components of the vector field $\textit{v}$ in the extra physical leg $p$. 
    Therefore, the number of parameter it retains is lower ($2.2\%$) than the the number of parameters needed to encode the three vector components with the stacked or interleaved encodings (7\%).
     }\label{fig:Concatenated_Ek_snap}
\end{figure}

For this comparison, we only consider the energy spectrum $E(k)$ as a metric. We plot the results in Fig.~\ref{fig:Concatenated_Ek_snap}. We only plot the concatenated TT built from individual stacked TTs, because we empirically do not observe any difference with respect to the concatenated TT built from individual interleaved ones. However, we compare $E(k)$ against both the stacked and interleaved encodings. We highlight the impressive compression given by the concatenated TT with respect to three individual TTs: indeed, the upper bound for the total number of parameters in the former case is $2\chi^2 (N+1)$, whereas in the latter one is $3 (2\chi^2 N)$.

For convenience, we plot the results only for $\chi = 2000$. In fact, we observe that the energy spectra differ from each other for smaller $\chi$s, and are practical equivalent for larger ones. For $\chi = 2000$, $E(k)$ in the inertial range is perfectly matched by the concatenated TT. In terms of compression, the concatenated encoding with $\chi = 2000$ reduces the number of total parameters to the $2.6\%$ of the full vector representation. Three individual TTs with the same $\chi$ would reduce it to the $7\%$.

This suggest that whenever a high bond dimension is needed to accurately capture the inter-scale correlations in the fluid, the concatenated encoding might become relevant to further reduce the number of total parameters without sacrificing accuracy with respect to the energy spectrum.

\section{TT-based solver} \label{app:QTT-based solver}

Here, we present the TT-based solver for the Navier-Stokes equations. We use this to simulate the time evolution of the turbulent snapshots from the DNS dataset, as discussed in Sec.~\ref{subsection:3D MPS-based solver}.
Our solver is an extension of the TT framework introduced in Ref.~\cite{peddinti}, with natural extensions from 2D to 3D.
The distinguishing feature is that the time evolution is performed completely within the TT representation. We achieve this by representing the discretized differential operators, appearing in Eqs.~\eqref{eq:NS}, in their corresponding TT form known as Matrix Product Operators (MPOs). For the finite difference operators used in this work, the corresponding MPOs are analytically constructed with low bond dimensions~\cite{kazeev2012low}.

In Table~\ref{table:complexity}, we report the asymptotic complexities for several steps of the TT framework. There, we include the complexities for two modes of evaluation of the TT solution along with retrieving the full-resolution solution. Since the memory cost of storing the full-resolution solution grows exponentially with $N$, these two modes, namely \textit{pixel sampling} and \textit{coarse-graining}, serve as an efficient alternative. These were also introduced in the proposed framework~\cite{peddinti}, to which we refer for detailed explanations. In this work, we can still afford to compute the full resolution solution. Hence, these two evaluation modes are not used in this work. However, once again, these evaluation modes will become indispensable at high resolutions when the full resolution is prohibited by the exponential memory requirements. For $N = 30$, the resulting vector is still small enough for us to evaluate the full vector from the compressed TT. We then compute the various measures, such as $E(k)$, from the full resolution solution.

\begin{table}
    \centering
    \begin{tabularx}{\columnwidth}{@{}XY@{}}
         \toprule
         \toprule
         \textbf{Algorithmic task} & \textbf{Time complexity}\\
         \toprule
         Divergence-free projection & $\mathcal{O}(N\,\rchi^6)$\\
         \midrule
         Euler time stepping & $\mathcal{O}(\rchi^6)$\\
         \midrule
         Coarse-grained evaluation & $\mathcal{O}(N\,\rchi^3)$\\
         \midrule
         Pixel sampling (per pixel) & $\mathcal{O}(\rchi^3)$\\
         \midrule
         SVD truncation & $\mathcal{O}(N\rchi^3)$\\
         \bottomrule
         \bottomrule
    \end{tabularx}
    \caption[Complexity for various steps of the framework.]{\textbf{Asymptotic time complexities of the main TT subroutines of the 3D solver.} We report the asymptotic worst-case time complexity of each subroutine of the TT solver. We report the theoretical scalings of the algorithms with respect to the number of TT tensors ($N$), and the bond dimension ($\rchi$).}
    \label{table:complexity}
\end{table}

\paragraph{Time stepping and numerical stability :}

For completeness, we discuss the details of the time stepping scheme, following the exposition in Ref.~\cite{peddinti}. We evolve the velocity fields within the divergence-free manifold using the Chorin's projection method~\cite{chorin1967numerical,chorin1968numerical}. The spatial discretization is given by the uniform grid and the differentiation is computed using the finite difference method. For the discretization of time, we implement an explicit Euler time-stepping scheme. Starting from Eqs.~(\ref{eq:NS}), this results in the following:

\begin{equation}
    \label{eq:euler}
    \frac{\boldsymbol{v}_{t+\Delta t} - \boldsymbol{v}_t}{\Delta t}+ (\boldsymbol{v}_t \cdot \nabla )\boldsymbol{v}_t = -\frac{1}{\rho} \nabla p_{t+\Delta t} + \nu \nabla^2 \boldsymbol{v}_t,
\end{equation}
along with the divergence-free condition:
\begin{equation}
    \nabla \cdot \boldsymbol{v}_{t+\Delta t} = 0 .
\end{equation}

Next, ignoring the pressure term in Eq.~\eqref{eq:euler}, we define an intermediate velocity field given by
\begin{equation}
    \label{eq:int_vel}
    \boldsymbol{v}_{t+\Delta t}^* = \boldsymbol{v}_t + (-(\boldsymbol{v}_t \cdot \nabla ) \boldsymbol{v}_t+ \nu\, \nabla^2 \boldsymbol{v}_t)\times\Delta t.
\end{equation}
However, the intermediate velocity does not satisfy the divergence-free condition.
We use the Helmholtz decomposition of $\boldsymbol{v}_{t+\Delta t}^*$ to define the solenoidal and irrotational components of the vector field. By definition, the solenoidal field has zero divergence at all points, which is indeed our objective.

Next, instead of finding the solenoidal component directly, we determine the irrotational component of the intermediate velocity. As stated in Chorin's projection method~\cite{chorin1967numerical}, this reduces to solving the Poisson equation for the pressure field:
\begin{equation}
    \nabla^2 p_{t+\Delta t} = \frac{\rho}{\Delta t} \nabla \cdot \boldsymbol{v}_{t+\Delta t}^* \ ,
\end{equation}
which we solve using a DMRG-based linear system solver~\cite{oseledets2012solution}.
Here, we remark that alternatives such as the TT-AMEn solver~\cite{Dolgov_Savostyanov_2014} could be used to further optimize the convergence and stability of the linear solver. We consider this enhancement out of scope for this work and leave it open for future work.

By subtracting the gradient of the pressure field, we find the solenoidal component of the velocity field, which completes the time evolution for one time step:
\begin{equation}
    \boldsymbol{v}_{t+\Delta t} = \boldsymbol{v}_{t+\Delta t}^* - \frac{\Delta t}{\rho} \nabla p_{t+\Delta t}.
\end{equation}

Moreover, we need to choose the time step size ($\Delta t$), which affects the stability of the time stepping. For a stable time evolution, we have to satisfy the Courant–Friedrichs–Lewy (CFL) condition~\cite{cfl_1928}. It states that for ($\frac{U\Delta t}{\Delta x} \leq 1$), where $U$ is the characteristic velocity, the information about the flow travels slower than the flow itself, ensuring stable numerical integration. We emphasize that this is only a necessary condition, not sufficient for the algorithm's stability.
In the 3D simulations reported in Sec.~\ref{subsection:3D MPS-based solver} we set $\Delta t = 2 \times 10^{-4}$. Other simulation parameters include $\nu = 6.7 \times 10^{-4}$ and $\mathrm{Re}_{\lambda} = 315$, in agreement with the dataset~\cite{Cardesa_2017}.

\paragraph{Bond dimension truncation :}
Each time evolution iteration involves several TT operations, such as element-wise multiplication (non-linear term in~\eqref{eq:NS_eq_a}), summation (Euler time step), DMRG-type algorithm (Projection step) and TT contractions with differential operators.
Each of them increases the bond dimension of resulting TT. Hence, we perform TT-rounding after each operation to a fixed bond dimension, for an efficient time evolution. We use the SVD-based truncation algorithm and limit the number of allowed singular values to fix the bond dimension.

\paragraph{DMRG-type solver for linear systems :}
As already discussed, in order to project the velocity fields onto the divergence-free manifold, we solve the resulting linear system using a DMRG-based algorithm. This task include two major components: tensor contractions and solution of the \textit{local} linear systems. For a fixed bond dimension, the time complexity is then estimated as the combined cost of tensor contractions needed to determine the local systems and the solution of the local linear systems. The tensor contractions are optimized in a way that the most information is reused from the previous DMRG sub-sweep, which scale as ($\mathcal{O}(N\rchi^4)$). The resulting local systems are of size $4\rchi^2\times 4\rchi^2$ and their exact solution scales as ($\mathcal{O}(\rchi^6)$). Thus, the resulting scaling for the projection step scales as in $\mathcal{O}(N\rchi^6)$.

Previous works~\cite{gourianov2022quantum} included variational approaches to tackle the linear system solution with a favorable worst-case complexity $\mathcal{O}(N\rchi^4)$. However, as already emphasized in~\cite{oseledets2012solution}, the exact solution is to be preferred when the bond dimensions are small enough to allow direct computation.

\paragraph{Immersed objects and masks :}
Moreover, in~\cite{peddinti}, a great deal was put into the idea of \textit{mask} that allows to simulate fluid flows around complex geometries, enforcing non-trivial boundary conditions by building the TTs associated to the immersed objects themselves.
However, in the context of this work, we are not interested in that construction as we look at a periodic cubic empty domain. Despite this, we remark that numerical simulations of turbulent fluids often involve non-trivial boundary conditions to make the turbulent behavior arise earlier in the dynamical evolution, or at lower Reynolds numbers. One famous example being grid turbulence. Therefore, the concept of \textit{mask} might become of practical relevance in future works when trying to simulate grid turbulence or, more in general, flows around complex geometries in 3D for high Reynolds numbers.

\clearpage

\end{document}